\documentclass[fleqn,usenatbib]{mnras}

\usepackage{newtxtext,newtxmath}

\usepackage[T1]{fontenc}
\usepackage{ae,aecompl}

\usepackage[normalem]{ulem}

\usepackage{graphicx}	
\usepackage{amsmath}	
\usepackage{import}
\usepackage{multirow}
\usepackage{booktabs}
\usepackage{caption}
\usepackage{subcaption}
\usepackage[ruled,vlined]{algorithm2e}
\usepackage{siunitx}
\usepackage{textcomp}
\usepackage{url}

\title[Closed Loop Predictive Control of AO Systems]{Closed Loop Predictive Control of Adaptive Optics Systems with Convolutional Neural Networks}

\author[Swanson et al.]{
Robin Swanson,$^{1, 2}$\thanks{E-mail: robin@cs.toronto.edu}
Masen Lamb,$^{2, 3}$
Carlos M. Correia,$^{4, 5}$
Suresh Sivanandam,$^{3, 2}$
\newauthor
Kiriakos Kutulakos$^{1}$
\\
$^{1}$Department of Computer Science, University of Toronto, 40 St. George Street, Toronto, Canada\\
$^{2}$Dunlap Institute for Astronomy and Astrophysics, University of Toronto, 50 St. George Street, Toronto, Canada\\
$^{3}$David A. Dunlap Department of Astronomy and Astrophysics, University of Toronto, 50 St. George Street, Toronto, Canada\\
$^{4}$Space ODT, Rua A. C. Monteiro, 65, 4050-014, Porto, Portugal\\
$^{5}$CENTRA, Faculdade de Engenharia da Universidade do Porto, Rua Dr. Roberto Frias s/n,
4200-465 Porto, Portugal}

\date{Accepted XXX. Received YYY; in original form ZZZ}

\pubyear{2015}

\begin{document}
\label{firstpage}
\pagerange{\pageref{firstpage}--\pageref{lastpage}}
\maketitle

\begin{abstract}
Predictive wavefront control is an important and rapidly developing field of adaptive optics (AO). Through the prediction of future wavefront effects, the inherent AO system servo-lag caused by the measurement, computation, and application of the wavefront correction can be significantly mitigated. This lag can impact the final delivered science image, including reduced strehl and contrast, and inhibits our ability to reliably use faint guidestars. We summarize here a novel method for training deep neural networks for predictive control based on an adversarial prior. Unlike previous methods in the literature, which have shown results based on previously generated data or for open-loop systems, we demonstrate our network's performance simulated in closed loop. Our models are able to both reduce effects induced by servo-lag and push the faint end of reliable control with natural guidestars, improving K-band Strehl performance compared to classical methods by over 55\% for 16th magnitude guide stars on an 8-meter telescope. We further show that LSTM based approaches may be better suited in high-contrast scenarios where servo-lag error is most pronounced, while traditional feed forward models are better suited for high noise scenarios. Finally, we discuss future strategies for implementing our system in real-time and on astronomical telescope systems.
\end{abstract}




\section{Introduction}

The frontiers of ground based optical and infrared astronomy have been pushed to exciting new regimes through the assistance of contemporary adaptive optics (AO) systems. While the impact of this technology in modern day astronomy is undeniably successful, there are several important limitations manifested from within these systems that inhibit their potential to achieve even better performance. A prime example of this arises in the field of high-contrast imaging (\cite{males2018ground}), where the potential contrast performance is heavily constrained due to servo-lag within the AO system (\cite{Correia_2017}). 

Outside of the field of high-contrast imaging, another largely limiting factor is with respect to sky coverage, which in turn is fundamentally related to errors from photon noise as measured by the AO system’s wavefront sensor (WFS). As the demands of astronomers push the need for increased sky coverage (particularly at high galactic latitudes), the need for techniques to increase this coverage becomes abundantly clear.

While servo-lag and photon noise arise from seemingly disparate origins, both present opportunities to apply new image processing techniques, such as deep learning, where both prediction and removal of noise from a series of images has successfully been demonstrated in other applications (\cite{claus2019videnn}, \cite{mathieu2015deep}). By leveraging these techniques, there is great potential to both increase the performance of existing AO systems by means of mitigating servo-lag (with minimal changes to the current AO pipeline), while also increasing the availability of guide stars and thereby improving sky coverage; important for science use cases for all AO systems (\cite{ellerbroek1998adaptive}).

In this paper, we present a method for training and employing neural network-based closed-loop AO controllers, which can improve both of these objectives, based entirely on wavefront slopes which are readily available from an AO systems. We show that these methods can be operated at a variety of guide star magnitudes, are robust over a wide range of seeing conditions, and greatly reduce both servo-lag and photon noise induced errors.

\begin{figure*}
    \centering
    \includegraphics[scale=0.55]{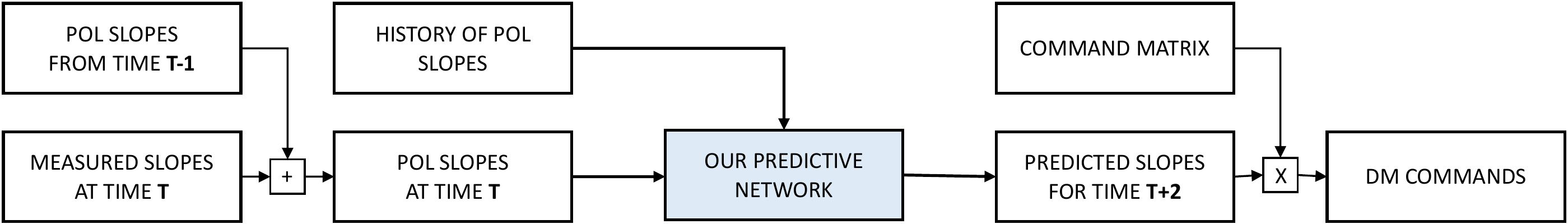}
    \caption{One iteration of our simulation when applying our predictive control algorithm. First, the POL slopes are calculated with the classical POL integrator. These and a small history of previous slopes are passed to our learned predictive networks. The output from our network corresponds to slopes two time steps ahead to mitigate servo-lag error. Finally, to generate the accompanying mirror commands we multiply the corrected slopes with the previously calibrated DM command matrix, resulting in the shape to be applied to the DM.}
    \label{fig:prediction_iteration}
\end{figure*}

\subsection{Neural Networks in Adaptive Optics}

Neural networks, and more recently deep convolutional neural networks (CNNs), are now well established as an effective and robust tool for many image processing problems~(\cite{lecun2015deep}). Their ability to learn complex and robust functions largely comes from the rich features that can be learned with a combination of using many 2D convolutions and having a very large set of training data to learn from. These methods have been adopted in many areas of scientific imaging, including the medical and astronomical sciences (e.g., ~\cite{ronneberger2015u, dieleman2015rotation}). However, while their ability to analyze, process, and improve astronomical image data has been well established, their ability to improve the instrumentation tools themselves is still in its infancy. 

One successful application of neural networks in astronomical instrumentation has been in the field of adaptive optics, where traditional densely-connected neural networks were applied to improve on the problem of off-axis anisoplanatism (\cite{gendron2011moao, osborn2012using}); this work showed a successful application of their model in an on-sky environment by training on wavefront sensor slope measurements. Other work applying more classical machine learning techniques to wavefront sensor processing (\cite{montera1996processing, And95spatio-temporalprediction} showed promising results over traditional methods, but were not applied to on-sky systems. However, because these non-convolutional networks are less capable of spatial and temporal reasoning, they are not able to leverage information across neighbouring slope measurements. They also have no ability to learn persistent patterns across wavefront measurements as each slope is processed individually. 

More recently, time-dependent neural networks have been applied to open-loop AO control (\cite{liu2020wavefront}), but were not integrated into a closed-loop integrator which is the typical mode of operation for most AO systems. Closed-loop neural network controllers can be difficult to train to be robust to system changes induced by its own output. In this work, we present a method for training a fully integrated closed-loop AO controller based on deep learning models which leverage both the temporal and spatial patterns inherently found in WFS slope measurements. We show that by constraining the network output with an adversarial prior, we enable higher performance compared to traditional integrators.

Finally, many upcoming instruments are exploring how to incorporate machine learning techniques into their systems (e.g., \cite{van2020robustness}), showing the growing demand for predictive control methods.

\subsection{Generative Adversarial Networks (GANs)}
\label{subsec:gans}

Generative Adversarial Networks~(GAN, \cite{goodfellow2014generative}) are a relatively recent development in deep learning which has greatly improved the realism of synthetically generated images. The key to their success is training two networks simultaneously: a generative network which learns to create realistic images, and a discriminator which learns do distinguish between real and fake images. As the two networks train, they compete against each other which helps the generator learn to create more realistic images that can fool the discriminator. 

Typically this technique is employed to generate new, previously unseen, images either from a random initialization~(\cite{goodfellow2014generative}) or conditioned on prior data~(\cite{liang2017dual}). Similarly, GANs have also found success in re-creating images based on the style of another~(\cite{zhu2017unpaired, Karras_2020_CVPR}), or to create more realistic training data for real-life systems~(\cite{shrivastava2017learning, james2019sim}). By treating the array of WFS slopes as an image we are able to leverage GANs in a novel way to facilitate training a closed-loop integrator (which is sensitive to input unlike those seen during training). Our discriminator acts as a prior on the output of the predictive network to be similar to the training data. Therefore when the loop is closed, and predictions from the network are fed back to the network as input, the input is still similar to the training data.

\subsection{Long-Short Term Memory Networks}
\label{subsec:lstms}

One area where CNNs prove less efficient is for temporally correlated data. While it is possible to incorporate time dependencies in our data by passing a large batch of data through the network at once, this does not explicitly encode any temporal structure for the model.

Recurrent neural networks, later improved by Long Short Term Networks (LSTM;~\cite{hochreiter1997long}), were created to solve this weakness by passing information from the last data point to the next one in time. Furthermore, in the case of LSTMs, the network creates and holds a hidden state at each layer in the network which is updated as new data is passed through the network. This enables the network to extract features and patterns over time which it finds relevant and produces better results as more data is passed through the network; this is accomplished by learning functions which choose relevant new data to add to the state at each time step, and which information can be forgotten. For a more detailed and illustrated description, see \cite{olah2015understanding}.

In cases where the data is both spatially and temporally correlated, such as videos or WFS sensor measurements, these methods can be further improved by using convolutional filters (\cite{xingjian2015convolutional}) such as those found in a typical CNN. These types of networks are aptly named convolutional LSTM networks and have been successfully used in a variety of cases such as video and motion prediction (\cite{finn2016unsupervised, lotter2016deep}); these networks are therefore a powerful tool for analyzing atmospheric phase perturbations which have strong spatial features and are highly temporally correlated.

\section{Deep Predictive Control}

\subsection{Network Architecture}
\label{subsec:netdescriptions}

\begin{figure}
    \centering
    \includegraphics[scale=0.6]{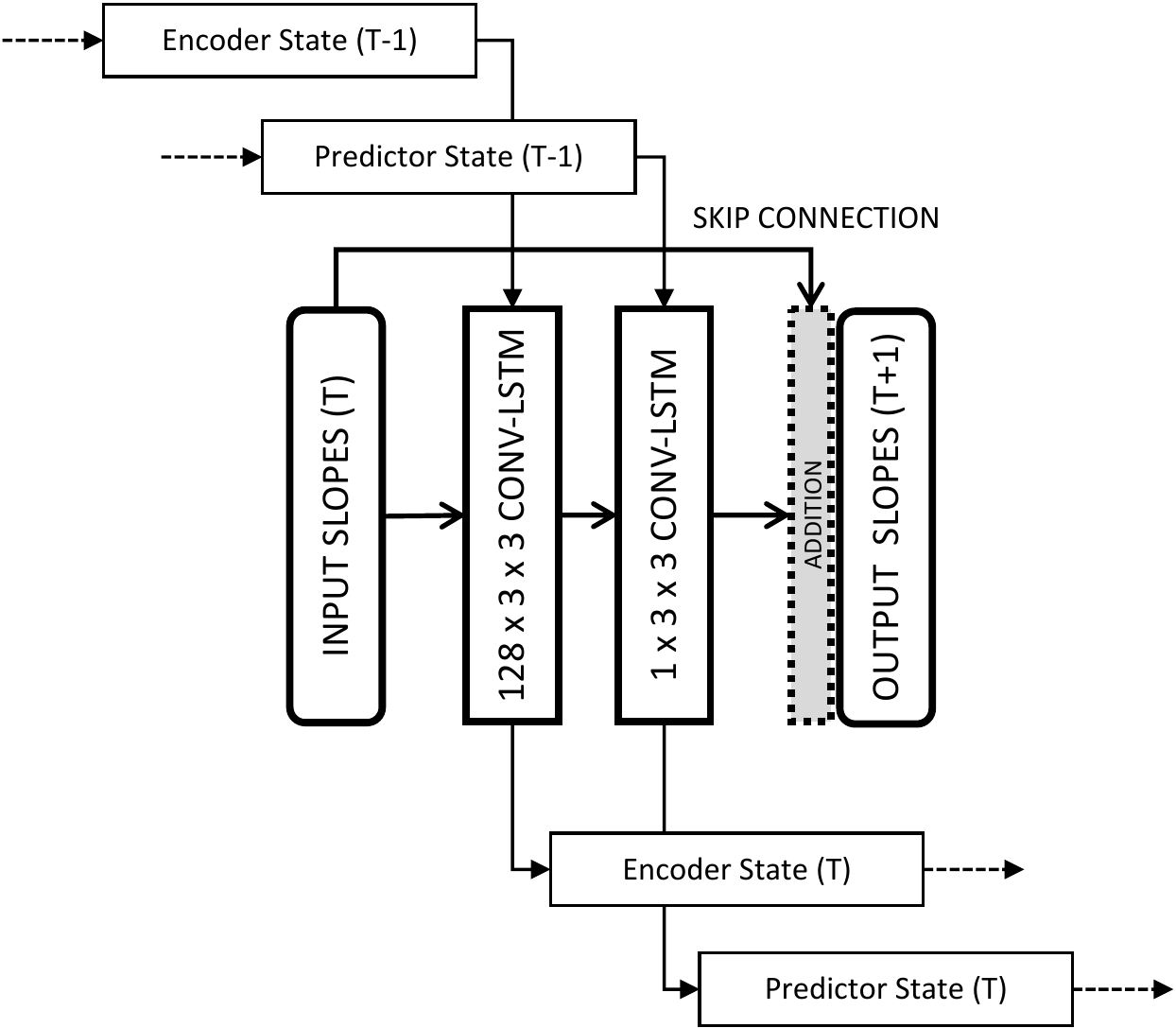}
    \caption{Overview of our predictive convolutional LSTM network. At each time step the current POL slopes and previous network states are passed into the network which predicts the upcoming slopes corresponding to those from two iterations forward in time. A single step of the LSTM network consists of two convolutional LSTM blocks, the encoder which takes a set of input slopes and convolves them with 128 pre-trained filters, outputting an encoded set of features and a state for the given iteration. Similarly, the predictor layer takes the encoded features from the encoder layer and predicts the next slopes based on that and the previous predictor state. In addition to each layers output, they also update their current state to be used in the next iteration.}
    \label{fig:lstmnetwork}
\end{figure}

\begin{figure*}
    \centering
    \includegraphics[scale=0.5]{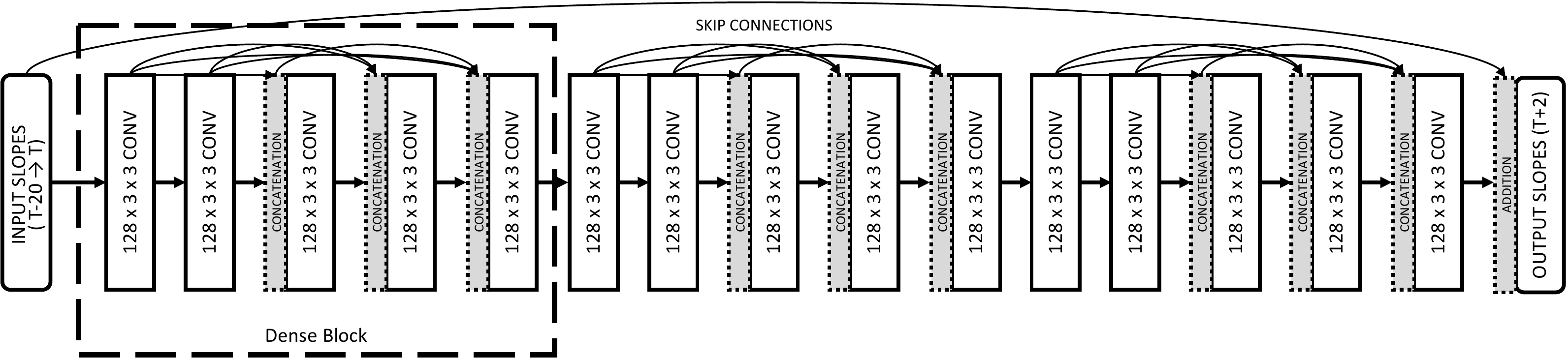}
    \caption{Overview of our densely connected predictive convolutional neural network. At each time step the latest 20 measured POL slopes are passed into the network which predicts the upcoming slopes corresponding to those from two iterations forward in time. The network consists of three densely connected blocks wherein each layer concatenates its output to the input channels of all future convolutional layers within its block. Each layer consists of $32$, $3\times3$ convolutional filters which are applied to all of its input channels after which a PreLU non-linear activation function is applied.}
    \label{fig:densenetwork}
\end{figure*}

\begin{figure}
    \centering
    \includegraphics[scale=0.50]{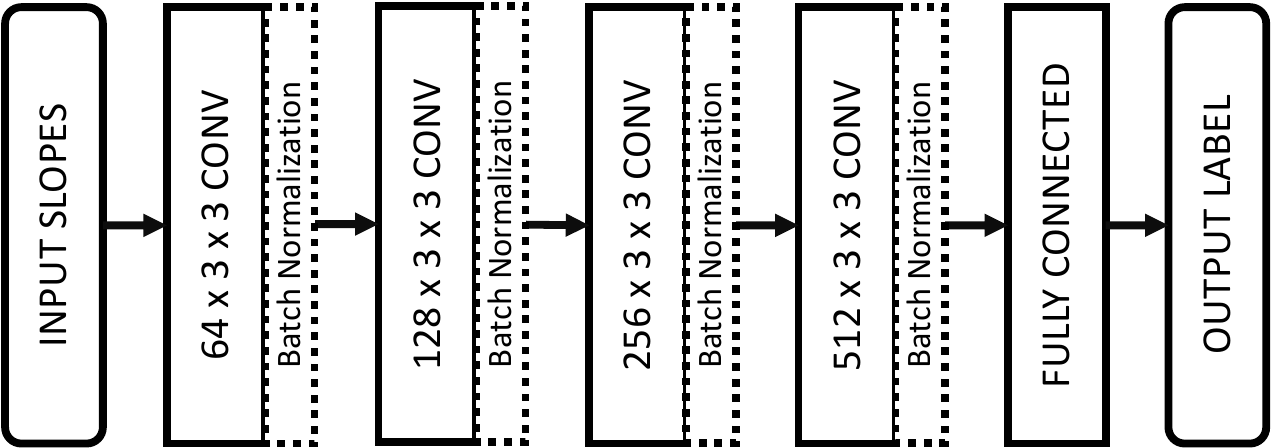}
    \caption{Overview of the discriminative network. During training this network attempts to label an input set of slopes as either "real" or "fake". Here "fake" refers to slopes output from our predictive network and "real" refers to those given as input to the network during training. The network consists of four $3\times3$ convolutional layers, each with an increasing number of filters. The final layer is a fully connected layer which maps the previous layers output to a single scalar value.}
    \label{fig:gannetwork}
\end{figure}

In this work, we consider two CNN architectures which incorporate time series data in different ways. Our goal is for these methods to function in any AO pipeline, as shown in Figure~\ref{fig:prediction_iteration}. We can then compare their performance against each other and the classical integrator to determine which neural network architectures are best suited for AO control. 

The first network uses convolutional LSTM layers and is shown in Figure~\ref{fig:lstmnetwork}. Here the slopes (in pseudo-open loop form; the motivation for which we describe in Section~\ref{subsec:netop}) from each time step are passed one by one through the network and at each iteration a state is saved and used to process subsequent data, thereby enforcing causality.

The second network is based on the densely connected convolutional neural network (CNN) architecture, as seen in Figure~\ref{fig:densenetwork}. Densely connected CNNs have been shown to learn meaningful convolutional features while maintaining a smaller number of learned parameters. In this case, the slopes from a fixed length of previous time steps are input as additional channels to the first layer of the neural network. Therefore, while the network does not have a strict sense of time, it can still recognize patterns and features across time by comparing each channel. 

We present results from such a network with three densely-connected blocks with five convolutional layers each. This design was chosen empirically, i.e., it contains the fewest number of blocks required to achieve comparable results to the LSTM model. However, in practice this number could be modified to meet computational cost and performance requirements.

Both networks include a skip connection that connects the input and output of the network. This reduces training time and network complexity~\cite{he2016deep}.

\subsection{Network Operation}
\label{subsec:netop}

Both networks were designed to exploit the natural spatial and temporal structure found in atmospheric phase profiles. For this reason we operate on pseudo-open-loop slopes (POL,~\cite{ellerbroek2003simulations}). Unlike residual slopes, that contain mostly high frequency noise, POL slopes are strongly correlated over time making prediction feasible. POL controllers are also regularly found in state of the art AO systems, unlike open-loop controllers.

We choose to do our inference on slope data due to them being the only available information from the atmospheric wavefront in typical AO telemetry. The large reduction in spatial dimensionality compared to a full resolution atmospheric wavefront also reduces computational complexity while preserving spatial and temporal relationships in the data. This assumption holds true so long as a bijection exists between a wavefront and its calculated slopes, which is typically true for this type of  single-conjugate AO system.

As input, both networks take a series of previous 2D slope vector fields. These are stacked and passed through the network as an $[N_x \times N_y \times 2T]$ matrix where $T$ is the number of time steps passed, and $N_x, N_y$ are the number of subapertures in the $x$ and $y$ direction, respectively. Once passed through the network, the slope predictions output from the network can be converted into deformable mirror (DM) commands with a pre-calibrated command matrix and applied directly onto the DM with no additional gain or processing.

Each layer of our networks contains a convolutional layer with $3\times3$ filters and use a PReLU activation function (\cite{he2015delving}). The non-linear PReLU function introduces a learned parameter $\alpha$ at each layer $i$ that allows the network to adapt its activation functions to the training data. Formally, it can be described as,

\begin{equation}
 f_i(y_i) = 
  \begin{cases} 
   y_i & \text{if } y_i \geq 0 \\
   \alpha_i y_i       & \text{if } y_i < 0
  \end{cases}
\end{equation}

where $y_i$ is an input channel of the $i^{th}$ layer of the network and $\alpha_i$ is the learned slope. 

\subsection{Adversarial Prior}

While we found empirically that these networks are sufficient for predicting slopes that accurately match the ground truth data (and thus could easily operate in open-loop), they did not perform well during closed-loop operation. This is due to statistical differences between the data distributions of the input, ground truth, and intermediate network output slopes, resulting in a network that performs well for a short amount of time in closed-loop before losing any improvements over classical methods. As the loop was closed using the latest output from our network, the new input to the network in subsequent iterations of the loop look less like the training data used to create the network, causing the network output to diverge.

One solution to this problem would be to reduce the number of learnable parameters of our network to avoid overfitting to the ground truth data. However, this in turn reduces the network's accuracy, resulting in very minor improvements over classical methods. As a solution to this problem, we include an additional discriminative network and loss function as a prior to encourage the network to output slopes which are close to the ground truth while statistically resembling the input data. In this way, the slopes output from our closed-loop integrator will be recognizable to the neural network as predictable input even if we are in entirely new simulation environments. 

The discriminative network takes a set of slopes as input and attempts to discern whether those slopes were part of the original training data set or a set of slopes produced by out network. We then train both the predictive and discriminative networks simultaneously, forcing our predictive network to both predict the future slopes while also fooling the discriminator. At the same time the discriminator is trained to better discern between the training slopes and the slopes output from our network. This alternating optimization continues until we have reached a saddle point in our combined loss function where the predictive network is as accurate as possible while the discriminator cannot reliably tell the two sets of input apart.

The discriminative network, as shown in Figure~\ref{fig:gannetwork}, is derived from the DCGAN model (\cite{radford2015unsupervised}) containing a simple series of strided convolutional layers, which take the input slopes and apply increasing numbers of convolutional filters at each layer while halving the spatial resolution at each step. This results in a final output from the convolutional layers of size $[512\times1]$ which is then fed into a fully connected layer that takes a learned, weighted sum of the outputs to produce a single output value between $0$ and $1$. Here an output of $0$ means the network believes the input slopes were produced by our predictive network while an output of $1$ means the network thinks it was from the original set of training data. 

While the discriminative network increases the training complexity, it is not required during inference after the weights of the network are fixed and therefore has no impact on the run-time of the final model.

\section{Training}

\subsection{Simulation Settings}
\label{sec:simulation} 

For both training and testing, we simulate the Gemini telescope; an 8-metre class telescope with a $16\times16$ lenslet array operating in POL at 800 Hz with two frames of servo-lag. We operated the AO with an R-band natural guide star, a three-layer atmosphere, and a K-band science camera imaging over a wide range of natural guide star (NGS) magnitudes. The read-out noise was purposefully kept to a negligible level to demonstrate our method's performance on photon noise alone. All simulations were implemented and run using the OOMAO adaptive optics simulation software package~\cite{conan2014object}.

To generate training data, 20000 independent simulations were run  for 500 loop iterations each. For better sampling of possible simulation settings, we randomly sample the Fried parameter ($r_0$) and wind speed from normal distributions, uniformly sample each wind direction from $(0, 2\pi]$, and uniformly sample the NGS magnitude between 8 and 16 for each simulation. This exposes our network to a wide variety of data during training thus helping it generalize for all simulation parameters and avoid overfitting to certain conditions. Please refer to Table~\ref{tab:simulationsettings} for further details.

At each simulation time step $t$, we save the current frame-delayed POL slopes, $\textbf{s}(t)$, used by the classical integrator to close the loop, as well as the current ground-truth ``best fit'' slopes $\textbf{s}^*(t)$, i.e., the slopes of the true atmosphere projected onto the DM. Given the current atmospheric phase $\Phi(t)$, as well as the command matrix $M$ and influence function $F$ of our calibrated DM, $\textbf{s}^*(t)$ is given by,

\begin{equation}
    \textbf{s}^*(t) = -\frac{1}{2}M^{-1}\left(F^{-1} \left( \Phi \left( t \right) \right) \right).
\end{equation}

The slopes $\textbf{s}$ and $\textbf{s}^*$ across all timesteps then represent our input and target pairs during training. To acquire a variety of conditions without considerable overlap we save five evenly spaced sets of 50 continuous time steps from each simulation. This gives us a total of 100,000 training pairs.

\begin{table}
\begin{tabular}{ccc} \toprule
\multicolumn{2}{l}{Simulation Parameters}         & Values                                       \\ \midrule
\multirow{10}{*}{Telescope}  & Diameter            & 8 m                                        \\
                            & Sampling Frequency  & 800 Hz                                     \\
                            & Frame Delay & 2 Frames                                 \\
                            & WFS Order              & $16\times16$                      \\
                            & WFS Readout Noise   & $\approx0 e^{-}$                                            \\
                            & DM Order                  & $17\times17$                \\
                            & Pupil Shape         & Gemini Pupil                               \\  
                            & NGS Band            & R                                            \\
                            & NGS Magnitude       & $\mathcal{U}(8, 16)$                                \\ 
                            & POL Gain            & $0.35$ \\ \midrule
\multirow{11}{*}{\vtop{\hbox{\strut Three Layer}\hbox{\strut Atmosphere}}} & $r_0$                & $\mathcal{N}(0.15, 0.02)$ cm                             \\ 
                            & Layers                & $3$                                         \\ \cmidrule(r){2-3}
                            & \multirow{3}{*}{Altitudes}                & $0$ km                          \\
                            &                                           & $4$ km \\
                            &                                           & $10$ km \\ \cmidrule[.1pt](r){2-3}
                            & \multirow{3}{*}{Fractional $r_0$}     & $0.70$ \\
                            &                                       & $0.25$ \\
                            &                                       & $0.05$ \\ \cmidrule[0.1pt](r){2-3}
                            & \multirow{3}{*}{Wind Speeds}  & $\mathcal{N}(5, 2.5)$ km/s \\
                            &                                       & $\mathcal{N}(10, 5)$ km/s \\
                            &                                       & $\mathcal{N}(25, 10)$ km/s \\ \cmidrule[0.1pt](r){2-3}
                            & Wind Directions     & $\mathcal{U}[0, 2\pi)$ rad                                 \\\midrule
\multirow{1}{*}{Science}    & Science Camera Band & K                                            \\ \bottomrule
\end{tabular}
\caption{Simulation parameters used for training our neural networks.} \label{tab:simulationsettings}
\end{table}

\subsection{Training}
\label{sec:training} 

During training, we take 20 sequential iterations of noisy input POL slopes $\textbf{s}$ from time $t$ to $t+20$ and similarly, the ground truth slopes $\textbf{s}^*$ for time $t+22$. This number of time steps was chosen due to it being the minimum sufficient number of time steps for training our networks (longer segments of time could be used at the cost of additional training time). The noisy slopes are concatenated and passed through the predictive network $\mathcal{P}$ which outputs a single set of slopes corresponding to time $t+22$. From this output we compute the data error term between $\mathcal{P}(\textbf{s})$ and $\textbf{s}^*$. For this we chose the $\ell_1$ vector norm loss which has been shown to have good convergence properties for CNNs and produce less noisy output~(\cite{zhao2015loss}). For a vector $\textbf{x}$ of length $n$ it is defined as,

\begin{equation}
    |\textbf{x}|_1 = \sum_{i=1}^n |\textbf{x}_i|.
\end{equation}

The full data loss term between our predicted and ground truth slopes is then given by,

\begin{equation}
    \mathcal{L}_{D} = ||\mathcal{P}(\textbf{s}) - \textbf{s}^*||_1.
\end{equation}

As noted in Section~\ref{subsec:gans}, we also include an adversarial loss to impose a prior on the network outputs to be as similar as possible to the training data. This takes the form of a binary Cross-Entropy loss function, penalizing outputs from the predictive network which are not labeled by the discriminative network $\mathcal{D}$ as the training data,

\begin{equation}
    \mathcal{L}_{A} = -log\left( \mathcal{D} \left( \mathcal{P} \left( \textbf{s} \right) \right) \right).
\end{equation}

At each training step the total loss of our predictive network is then,

\begin{equation}
    \mathcal{L}_{P} = \mathcal{L}_D + \gamma \mathcal{L}_A = ||\mathcal{P}(\textbf{s}) - \textbf{s}^*||_1 - \gamma log\left( \mathcal{D} \left( \mathcal{P} \left( \textbf{s} \right) \right) \right)
\end{equation}

where $\gamma$ weighs the importance of the two loss terms. By increasing the value of $\gamma$, we put higher weight on the adversarial loss and, in turn, decrease the network's ability to perfectly recreate the ground truth data. The weight is therefore required to balance between the two terms and its value can only be determined empirically by using the resulting method in closed loop. Based on the results of our experiments, we find that a value of $\gamma = 5$ works well and is used throughout this work.

Simultaneously with training the predictive network $\mathcal{P}$, we update the discriminative network $\mathcal{D}$ at each step based on the current trainable weights $\theta_\mathcal{P}$ of network $\mathcal{P}$. Similar to $\mathcal{L}_D$, $\mathcal{D}$ must maximize the likelihood of labeling both the input training data and the output of $\mathcal{P}$ correctly. This can be succinctly described by,

\begin{equation}
    \min_{\theta_\mathcal{P}} \max_{\theta_\mathcal{D}} \mathbf{E}\left[ log\left( \mathcal{P}_{\theta_\mathcal{P}} \left( \textbf{s} \right) \right) \right] + 
    \mathbf{E}\left[ log\left( 1 - \mathcal{D}_{\theta_\mathcal{D}} \left( \textbf{s} \right) \right)\right].
\end{equation}

Our network was implemented and trained using the Tensorflow  machine learning software package (\cite{tensorflow2015-whitepaper}). All optimizations were performed with a batch size of 32 using the Adam optimizer (\cite{kingma2014adam}) with an initial learning rate of $1e^{-4}$. The Adam optimizer is widely regarded as the current best method for optimizing deep neural networks, due in large part to its ability to tune the learning rate for each variable individually.

\subsection{Effect of Adversarial Prior}
\label{subsec:prioreffect}

\begin{figure}
    \centering
    \includegraphics[width=0.465\textwidth]{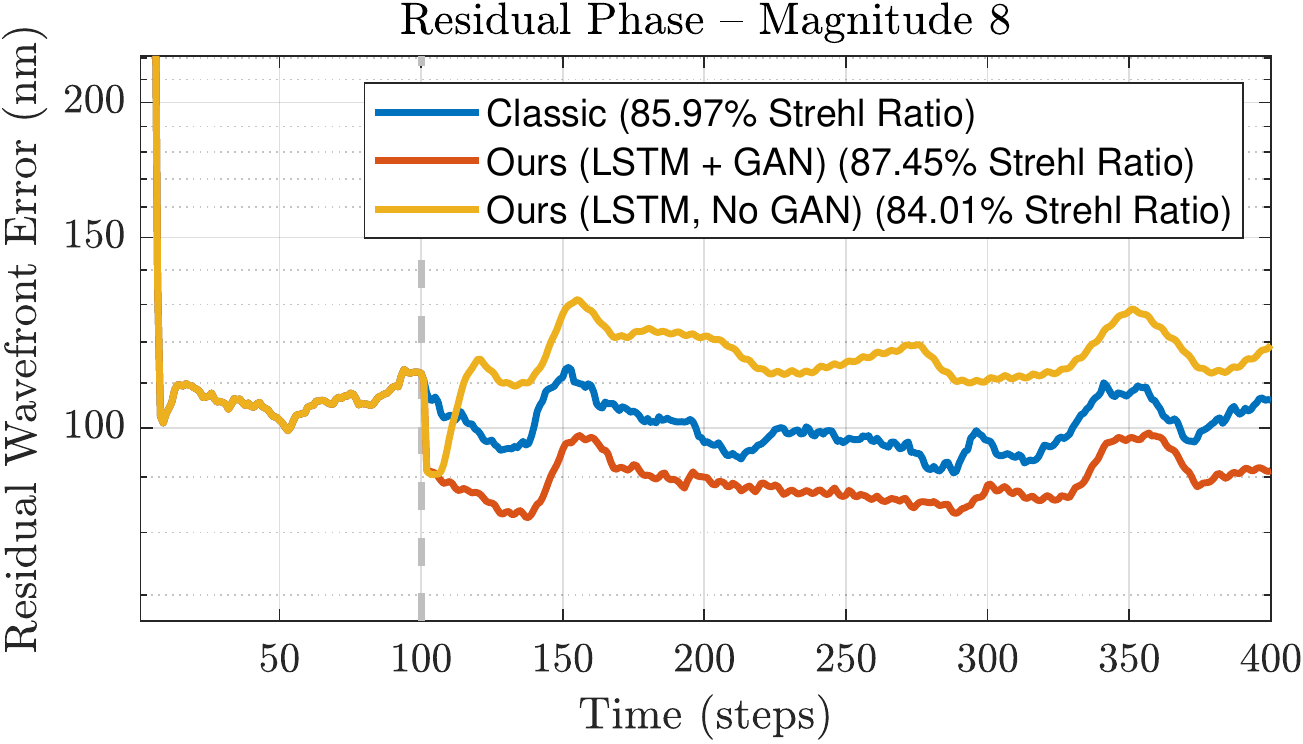}
    \caption{Closed loop RMS wavefront error performance of our LSTM network with and without the GAN prior compared to a classical integrator. Without the GAN prior, our network is unable to sustain its performance over many iterations of the simulation. Once trained with the adversarial training method, our network is more robust to inputs that have been influenced in its own input, enabling our controller to sustain its performance improvements over long periods of time.}
    \label{fig:example-openloop}
\end{figure}

As previously described, we use an additional adversarial prior to train our generative networks. Without this prior, we found that the network made excellent predictions in an open loop configuration but was prone to diverge or revert to classical integrator performance when running in closed loop \citep{swanson2018wavefront}. Typical performance for these networks can be seen in Figure~\ref{fig:example-openloop} which shows a great improvement over the classical predictions in open loop, but only for a short period after the network output has been used to close the loop.

We hypothesize that this is due to statistical differences between the input training data (i.e., noisy, POL slopes), and the expected ground truth values (ideal, noiseless slopes) used to train the network. Once the loop is closed, the network output is combined with the residual slopes calcualted at the next time step and used as input to the network. However, because the network is never trained on slopes similar to those it is trained to predict, it is unable to perform as expected. By including an adversarial we have found the network output appears more statistically similar to the training data, thus greatly mitigating the aforementioned closed loop effect. In this way, after the loop is closed with the neural network, the output slopes will still by recognized as valid input by the neural network at the next step and continue generating more accurate slopes.

Although other solutions may exist, we have found this approach works very well under all simulation settings and incurs only a small increase in training time. Furthermore, because it depends only on POL slopes, it will likely be adaptable to real-world scenarios where this is the only training information available.

\section{Results}

\begin{figure}
    \centering
    \includegraphics[width=0.465\textwidth]{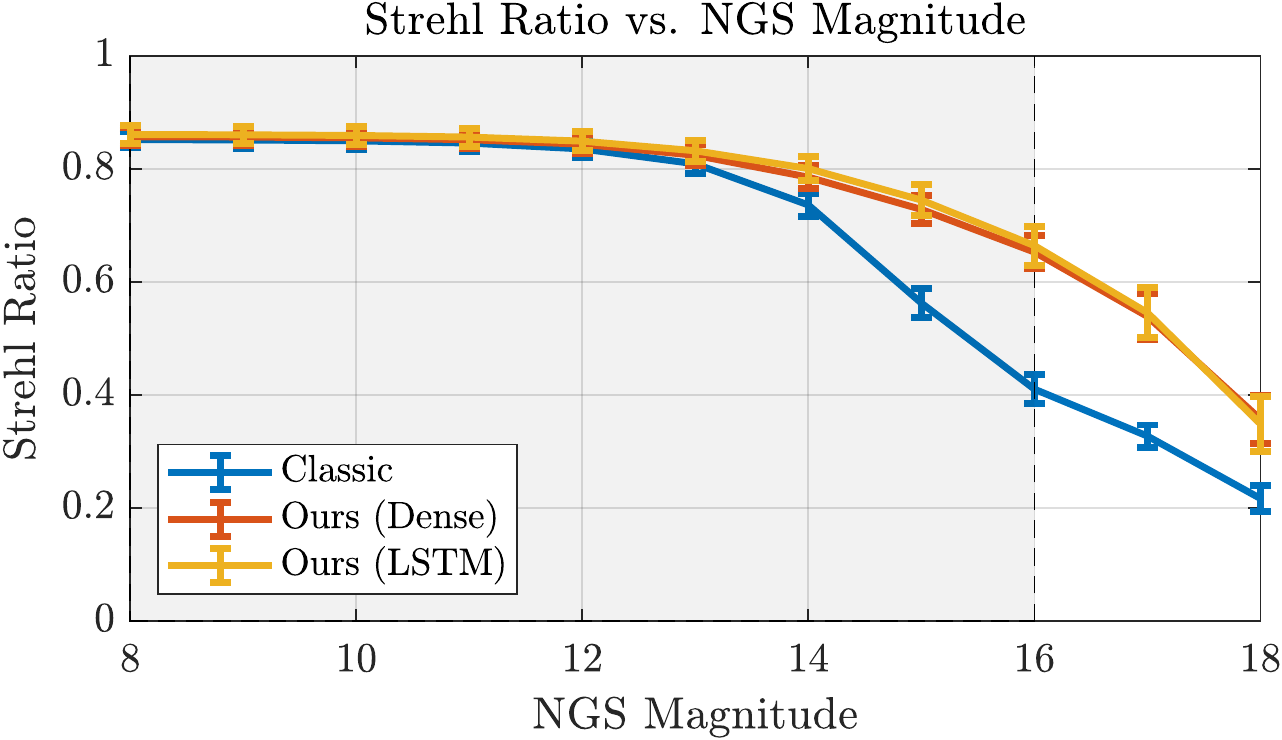}
    \caption{Closed loop performance comparison as a function of NGS magnitude. Our networks show considerable performance increase for faint NGS where photon noise dominates. Even under ideal NGS conditions our networks increase Strehl performance by reducing servo-lag induced errors. Our networks are also robust even to NGS magnitudes outside of the training data (highlighted in gray).}
    \label{fig:results_aggregate_strehl}
\end{figure}

\begin{figure}
    \centering
    \includegraphics[width=0.465\textwidth]{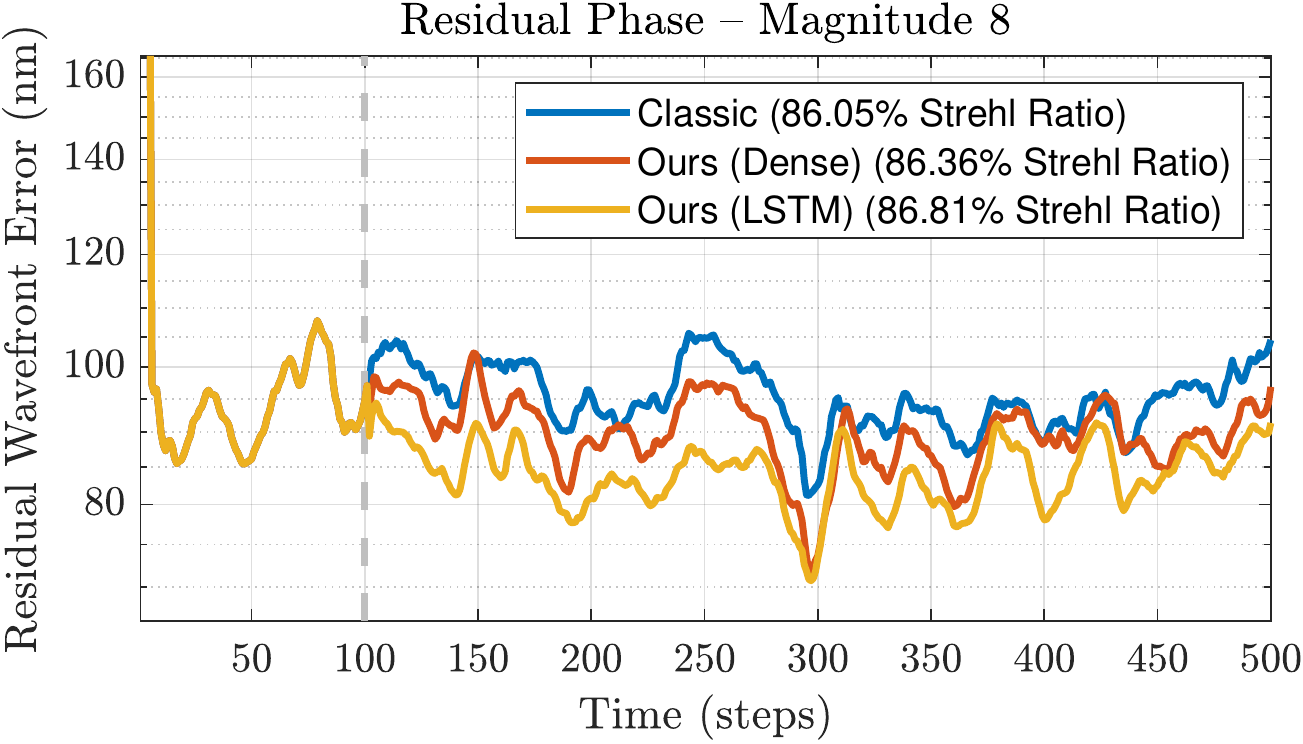}
    \\
    \vspace{2.5mm}
    \includegraphics[width=0.465\textwidth]{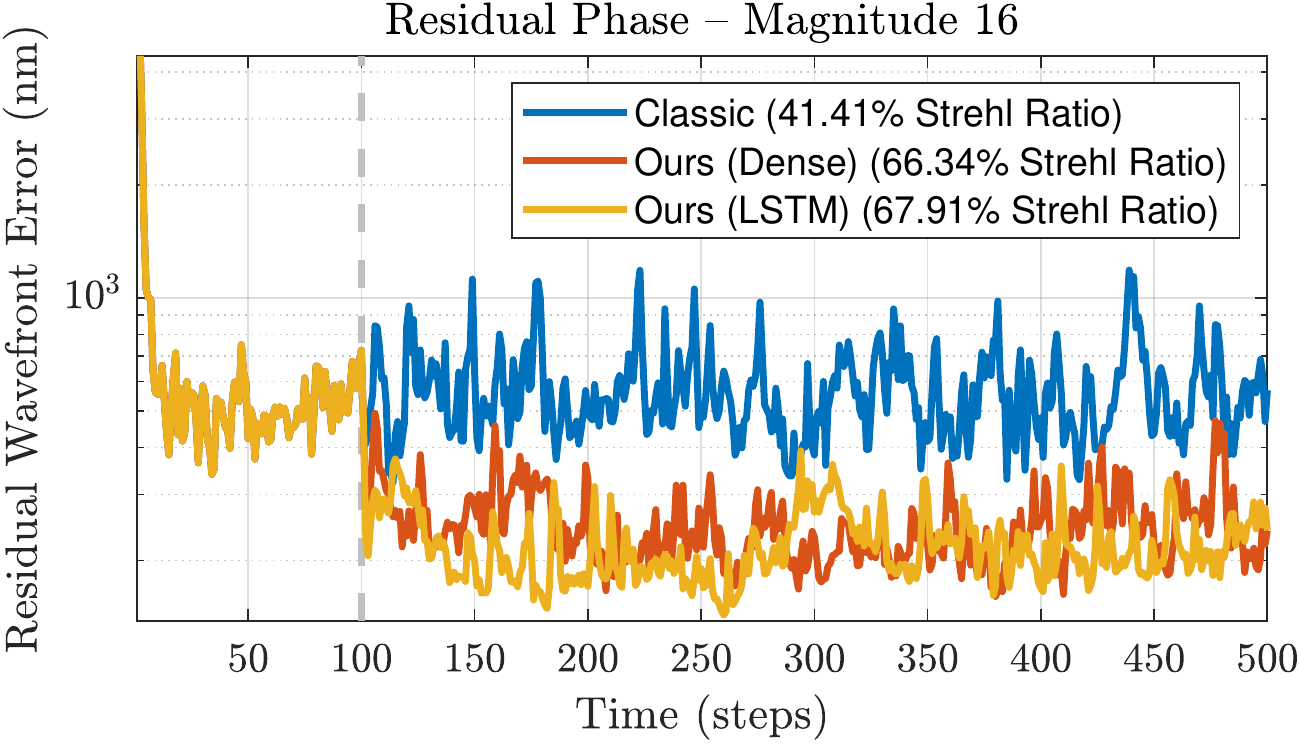}
    \caption{\textbf{Residual Wavefront Analysis:} Example residual RMS wavefront error measurements over the course of a simulation. We compare our predictive networks with a classical integrator for a magnitude 8 (top) and magnitude 16 (bottom) NGS. Our networks begin to predict after 100 iterations (indicated with a dashed veritcal line) and immediately shows improvements over the classical method. Strehl ratio values are included in the legend.}
    \label{fig:results_single_run}
\end{figure}

\subsection{Network Testing}
\label{sec:testing}

After training, we tested our network's performance by using their output to close the AO loop in previously unseen simulations. Each testing simulation was randomly initialized with the same range of parameter values used for the training simulations. Because our network was trained on converged, closed loop data, we use the classical POL integrator for the first 100 time steps before switching to our network to close the loop. To compare the performance of our networks with the classical method, we ran the same simulations for all methods, initialized with identical simulation parameters. In this way we can directly compare their performance for each simulation in addition to the aggregate performance across all simulations. Furthermore, for all experiments we show cases where the simulation parameters go beyond those used for training. These results reinforce our network's ability to generalize to unseen conditions.

\begin{figure}
    \centering
    \includegraphics[width=0.465\textwidth]{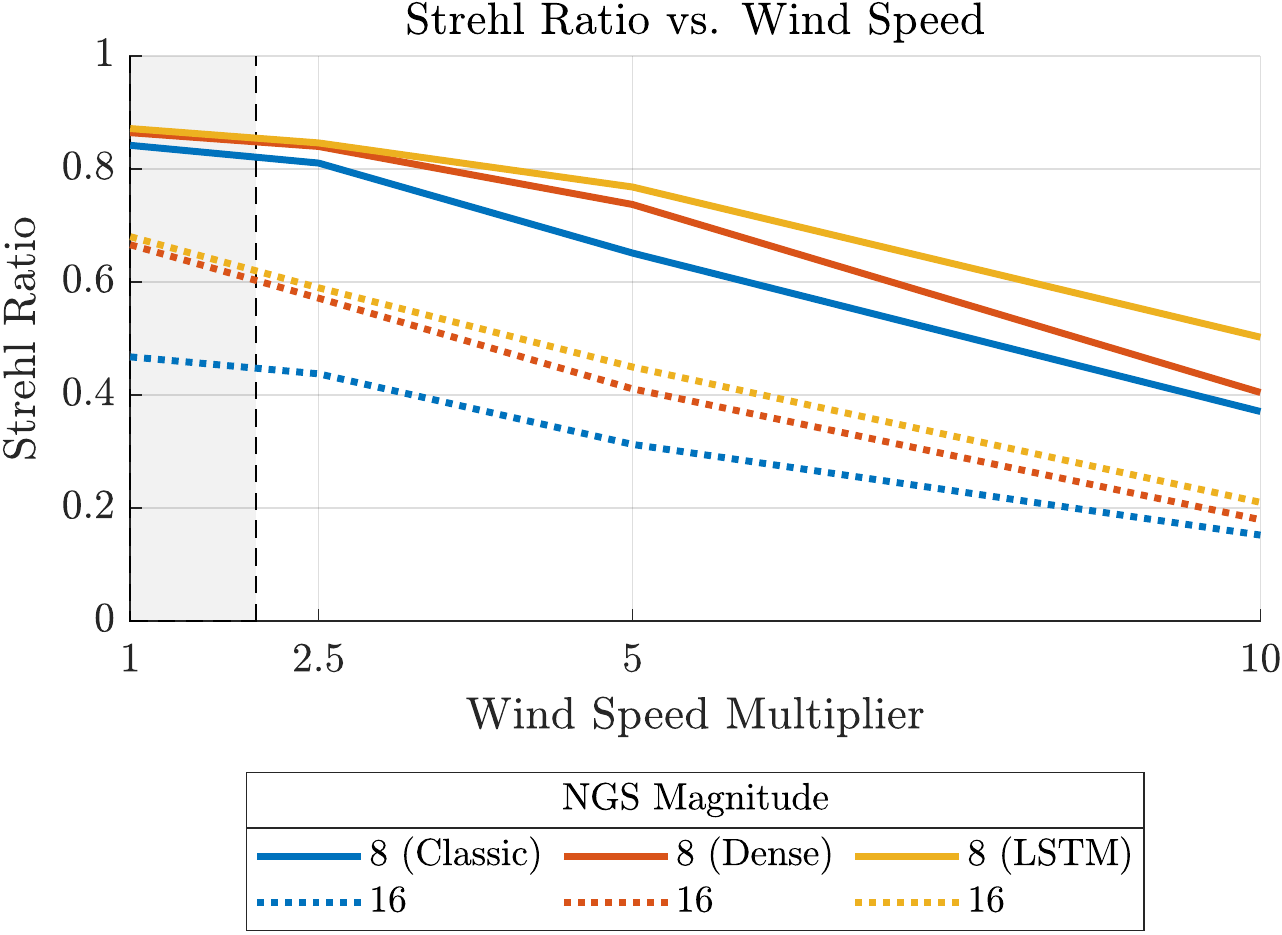}
    \caption{Closed loop performance comparison as a function of wind speed. While holding all other simulation variables constant, we mulitply a base windspeed for the three layer atmosphere of $[5, 10, 25]$ km/s by a factor of $[1, 2.5, 5, 10]$ to evaluate how well the three methods perform under increasing wind speeds for NGS magnitudes of 8 and 16. By leveraging the history of previous slope information our networks are able to maintain their performance well beyond the operating range of the classical integrator. They also perform well for conditions well outside the training data (highlighted in gray).}
    \label{fig:strehl_wind}
\end{figure}

\begin{figure}
    \centering
    \includegraphics[width=0.465\textwidth]{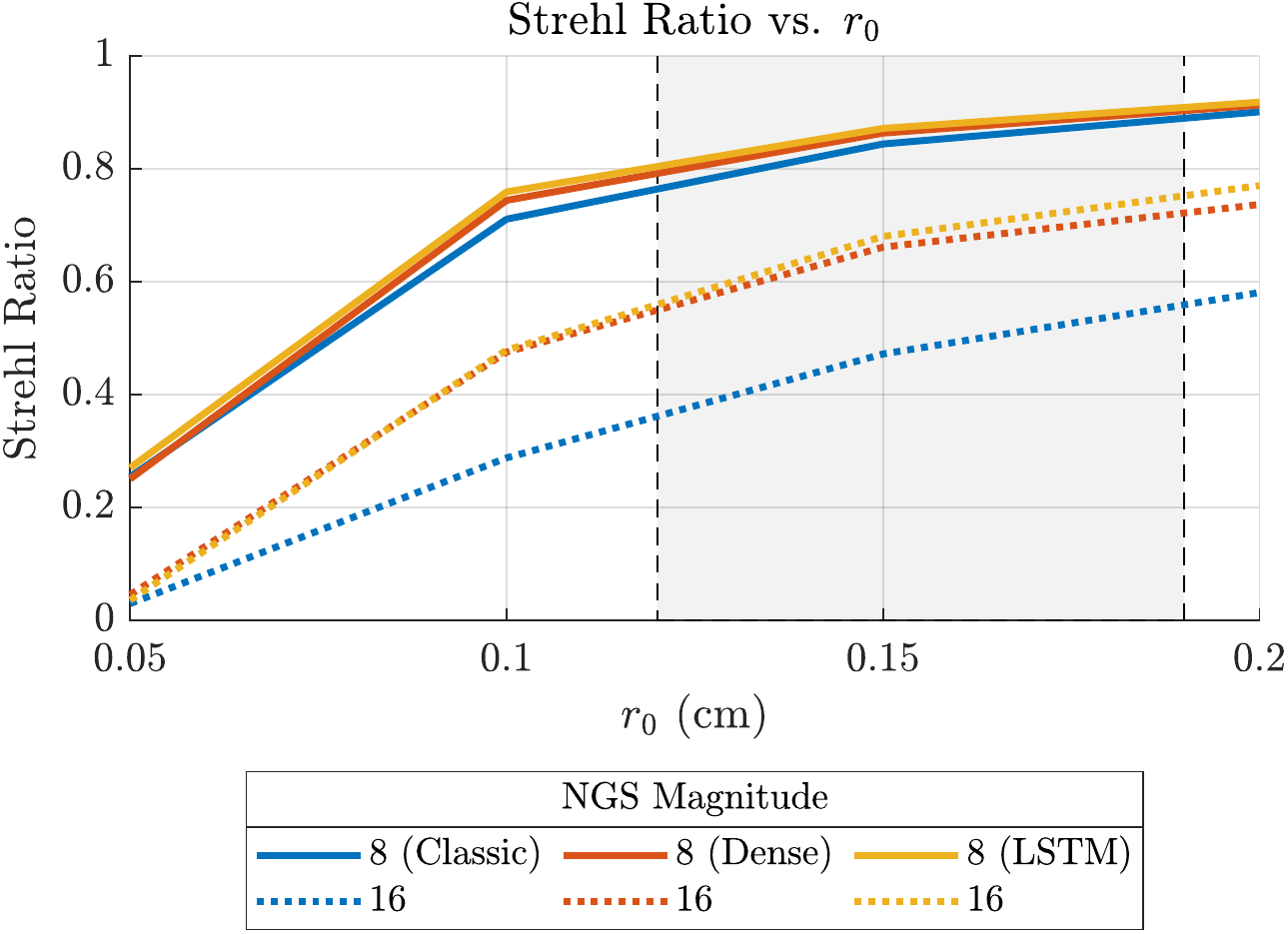}
    \caption{Closed loop performance comparison as a function of $r_0$. While holding all other variables constant, we vary the seeing conditions of the simulation by setting the $r_0$ value between $0.05$ and $0.2$cm. As the seeing conditions deteriorate, our networks are able to maintain their performance gains outside of typical operating ranges of $r_0$. They continue to perform well outside the ranges used to train the networks (highlighted in gray), showing their robustness to unseen conditions.}
    \label{fig:strehl_r0}
\end{figure}

\begin{figure}
    \centering
    \includegraphics[width=0.465\textwidth]{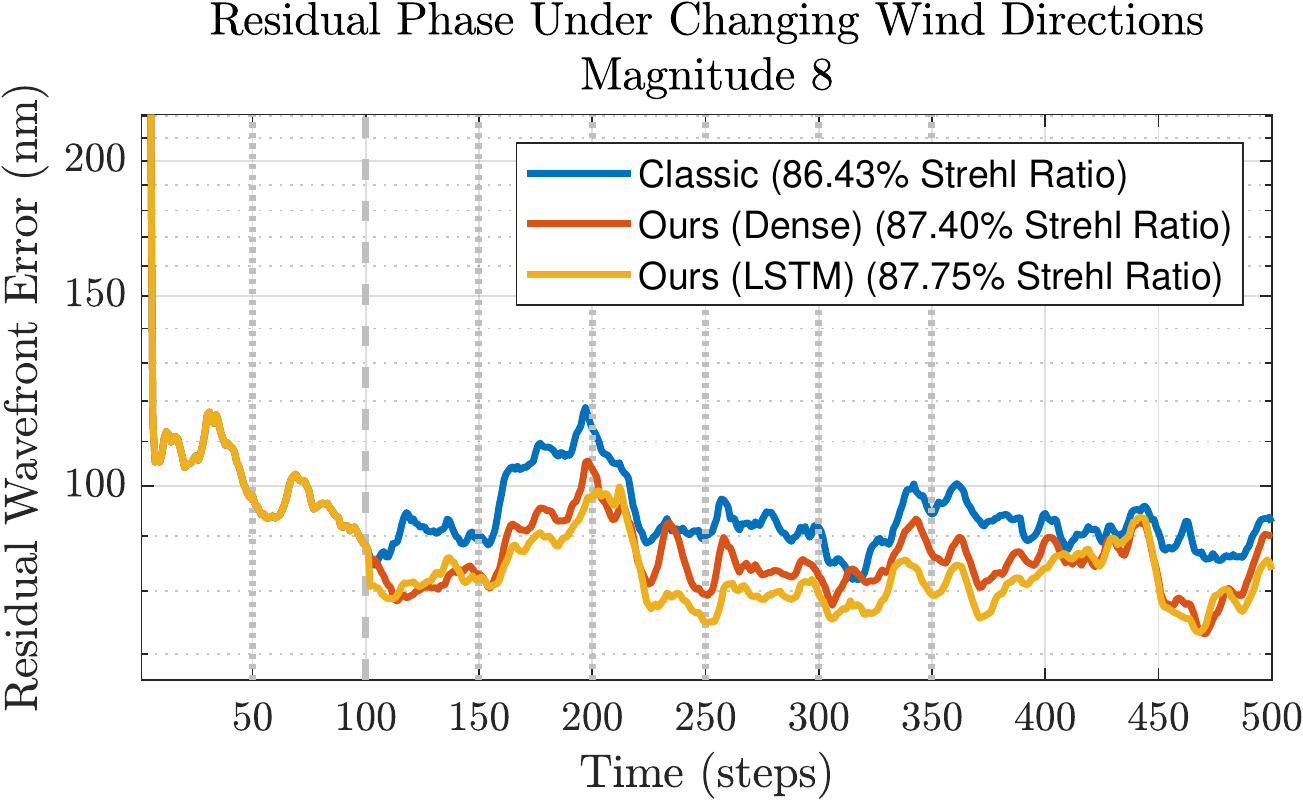}
    \\
    \vspace{2.5mm}
    \includegraphics[width=0.465\textwidth]{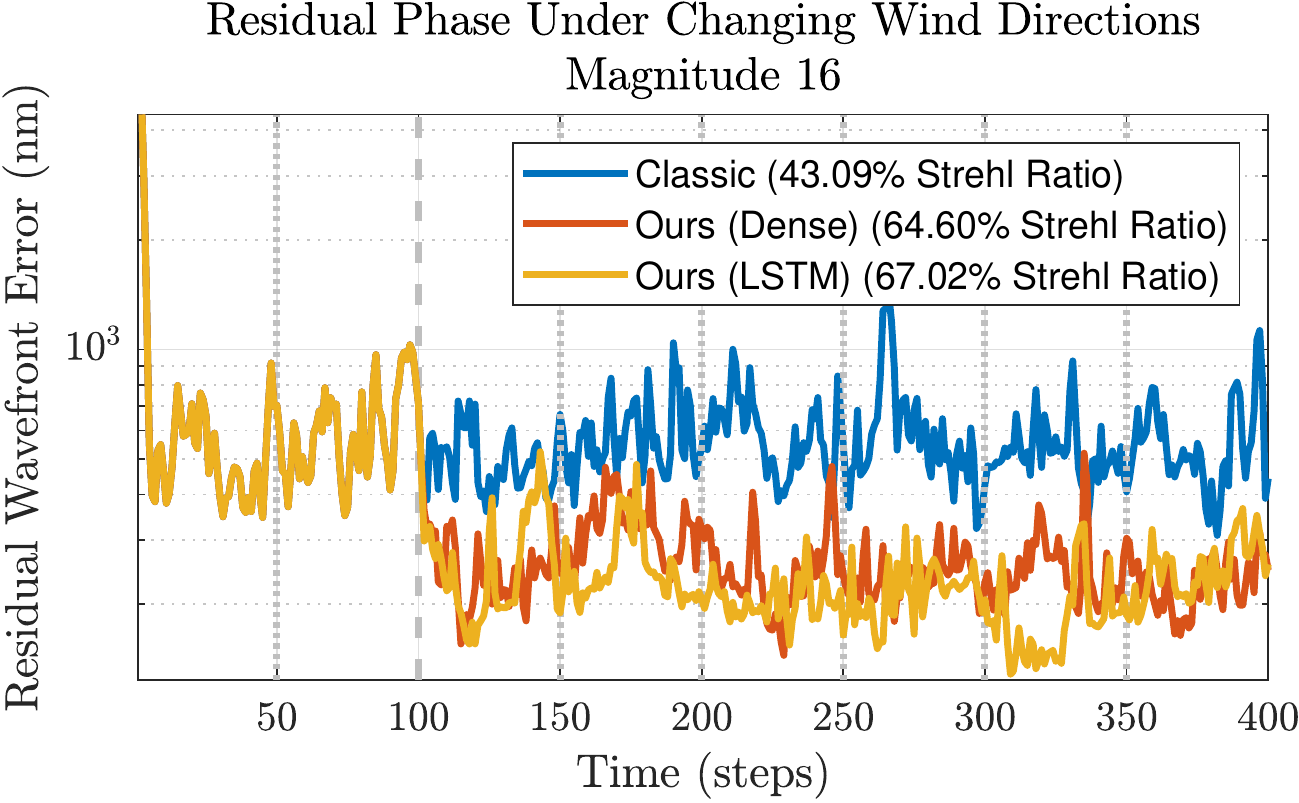}
    \caption{\textbf{Residual Wavefront Analysis Under Quickly Changing Wind Conditions:} Example residual RMS wavefront error measurements over the course of a simulation where the wind conditions change every 50 iterations (shown in dashed grey lines). We compare our predictive networks with a classical integrator for a magnitude 8 (top) and magnitude 16 (bottom) NGS. Despite being trained on constant wind directions, our models are able to take changing conditions into account. Strehl ratio values are included in the legend.}
    \label{fig:results_wind_change_run}
\end{figure}

\begin{figure}
    \centering
    \includegraphics[width=0.465\textwidth]{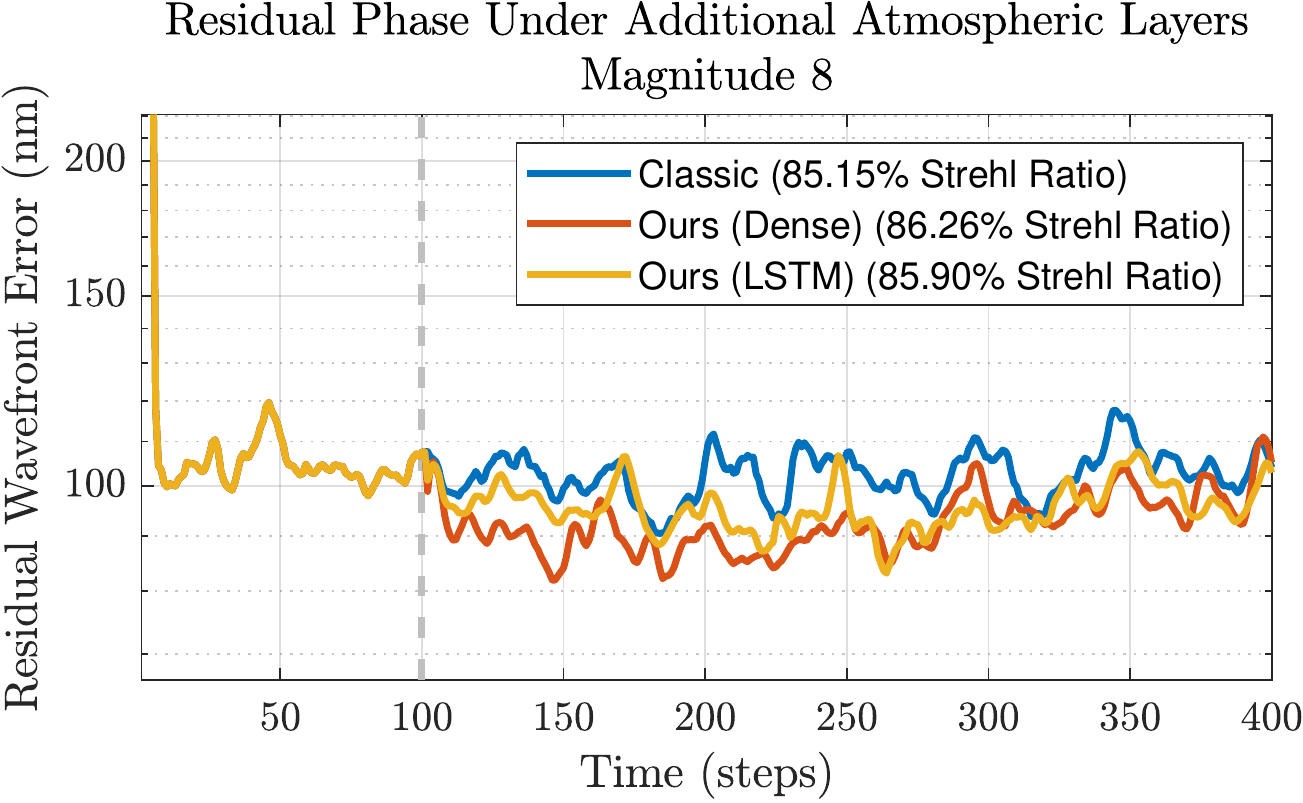}
    \\
    \vspace{2.5mm}
    \includegraphics[width=0.465\textwidth]{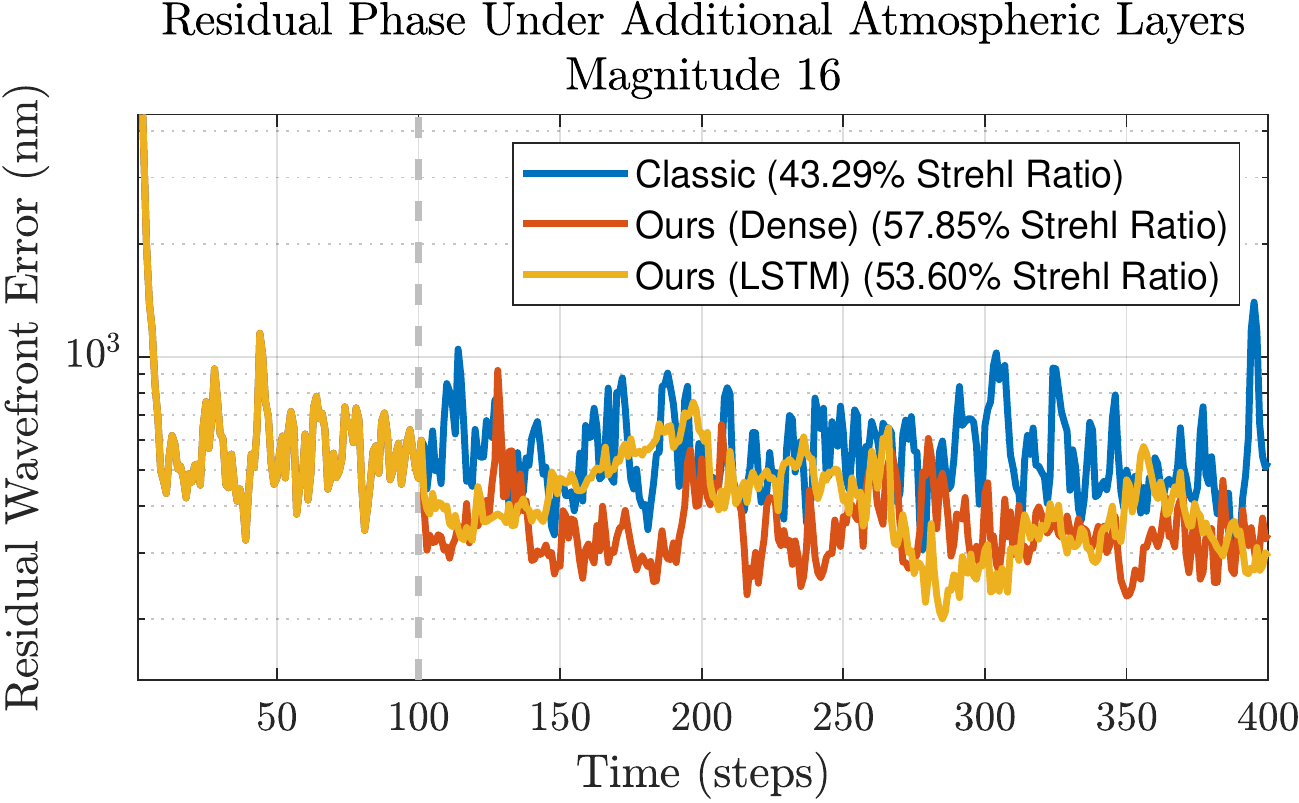}
    \caption{\textbf{Residual Wavefront Analysis Under Additional Layers of Atmosphere:} Example residual RMS wavefront error measurements over the course of a simulation where the testing conditions have many more layers of atmosphere than the training conditions. We compare our predictive networks with a classical integrator for a magnitude 8 (top) and magnitude 16 (bottom) NGS. While our networks were trained only with three layers of atmosphere, the dense network generalize very well to increased atmosphere complexity while the LSTM network shows more moderate generalization. Strehl ratio values are included in the legend.}
    \label{fig:results_more_atm}
\end{figure}

\begin{figure}
    \centering
    \includegraphics[width=0.465\textwidth]{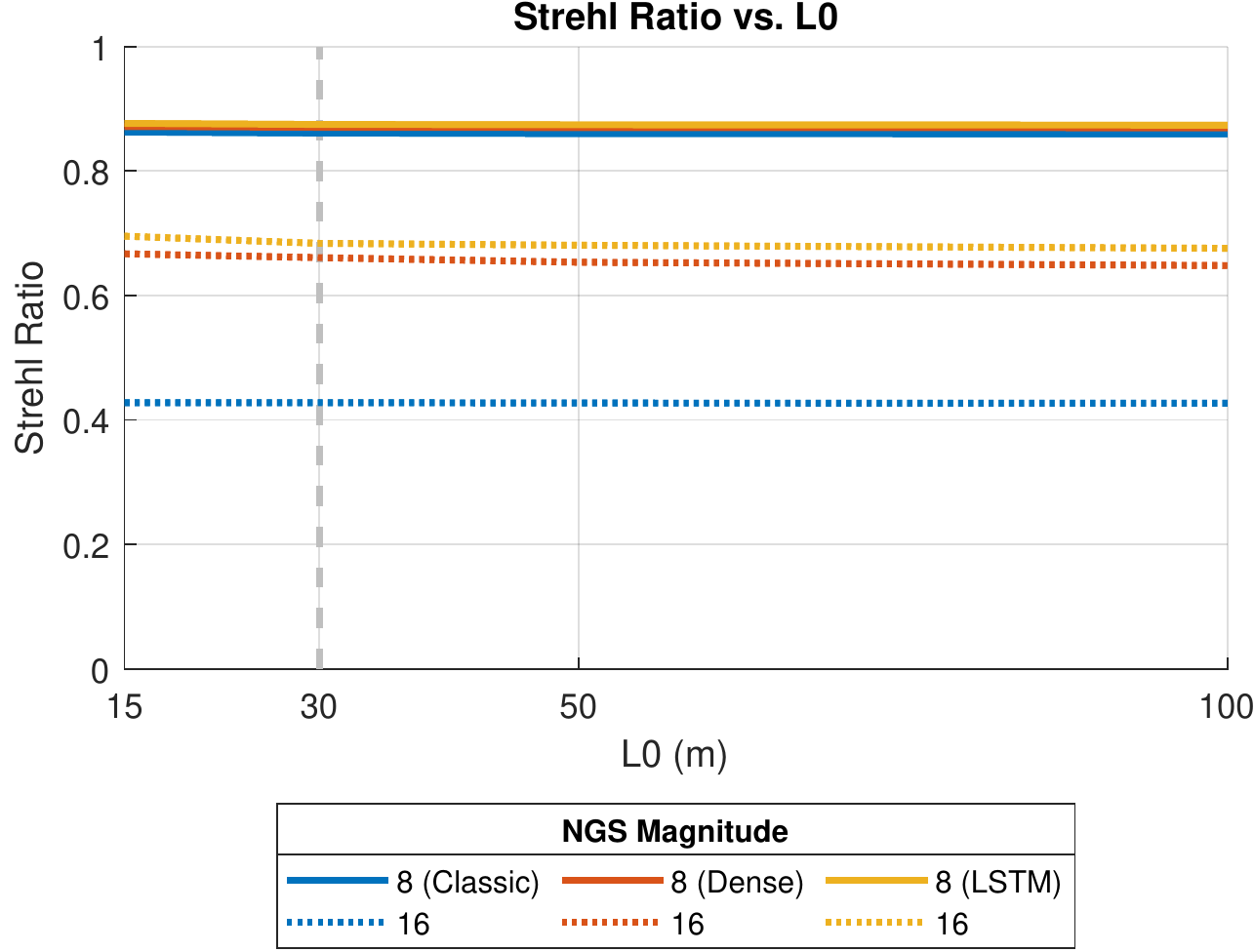}
    \caption{Closed loop performance comparison as a function of outer scale (L0). While holding all other simulation variables constant, we vary the outer scale of the atmosphere. Although our models were trained only with an L0 value of 30m (shown with a dashed grey line), all three methods see only a small change in performance as the outer scale changes.}
    \label{fig:results_change_L0}
\end{figure}

\subsection{Results}

In Figure~\ref{fig:results_aggregate_strehl} we show the average K-band Strehl ratio performance over 25 simulations at each NGS magnitude for our methods and the classical POL integrator (as reported by OOMAO). Our methods clearly improve the overall Strehl ratio across all star magnitudes. Not only at the faint end, where noise dominates and its effects are most significant, but also for bright sources where servo-lag and aliasing are the largest sources of non-fitting related error.

Figure~\ref{fig:results_single_run} shows the the residual wavefront error over time for a magnitude 8 and 16 NGS simulation. Our networks immediately improve the residual wavefront error after the necessary burn-in period for the closed loop integrator to reach a relatively steady-state (indicated with a dashed vertical line). These improvements are further increased as the seeing conditions become worse by either increasing wind speeds or decreasing $r_0$ of the system. 

Figure~\ref{fig:strehl_wind} compares our method with the classical integrator as the wind speed increases for all three atmospheric layers while $r_0$ is held constant. For all NGS magnitudes and windspeeds, our networks show improvements over the classical method, with the performance gap increasing as the winds grow stronger. For high wind speeds, our networks improve the Strehl ratio performance by the same factor as a decrease of 2 NGS magnitudes. Similarly, Figure~\ref{fig:strehl_r0} shows the performance of the two networks as the $r_0$ varies when wind speed and direction are held constant. Again, we see that at nearly ideal seeing conditions our networks show some strehl ratio improvements with the performance gap increasing as the seeing becomes worse at lower values of $r_0$ within the range of the training data.

\begin{figure*}
\captionsetup[subfigure]{labelformat=empty}
     \centering
      \begin{subfigure}[b]{0.32\textwidth}
         \centering         
         \caption{Classical PSD (85.7\% Strehl Ratio)} \vspace{-1.5mm}
         \includegraphics[width=\textwidth]{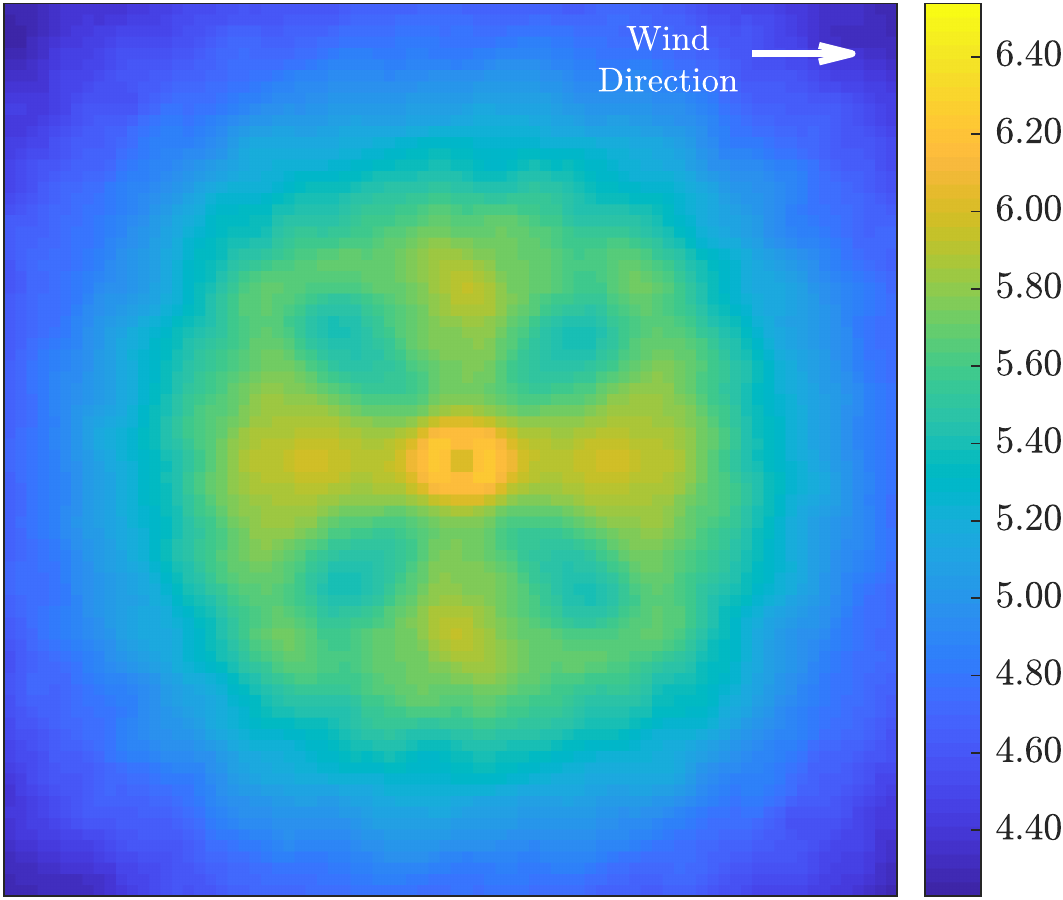}
     \end{subfigure}
     ~
     \begin{subfigure}[b]{0.32\textwidth}
         \centering
         \caption{Dense PSD (85.8\% Strehl Ratio)} \vspace{-1.5mm}
         \includegraphics[width=\textwidth]{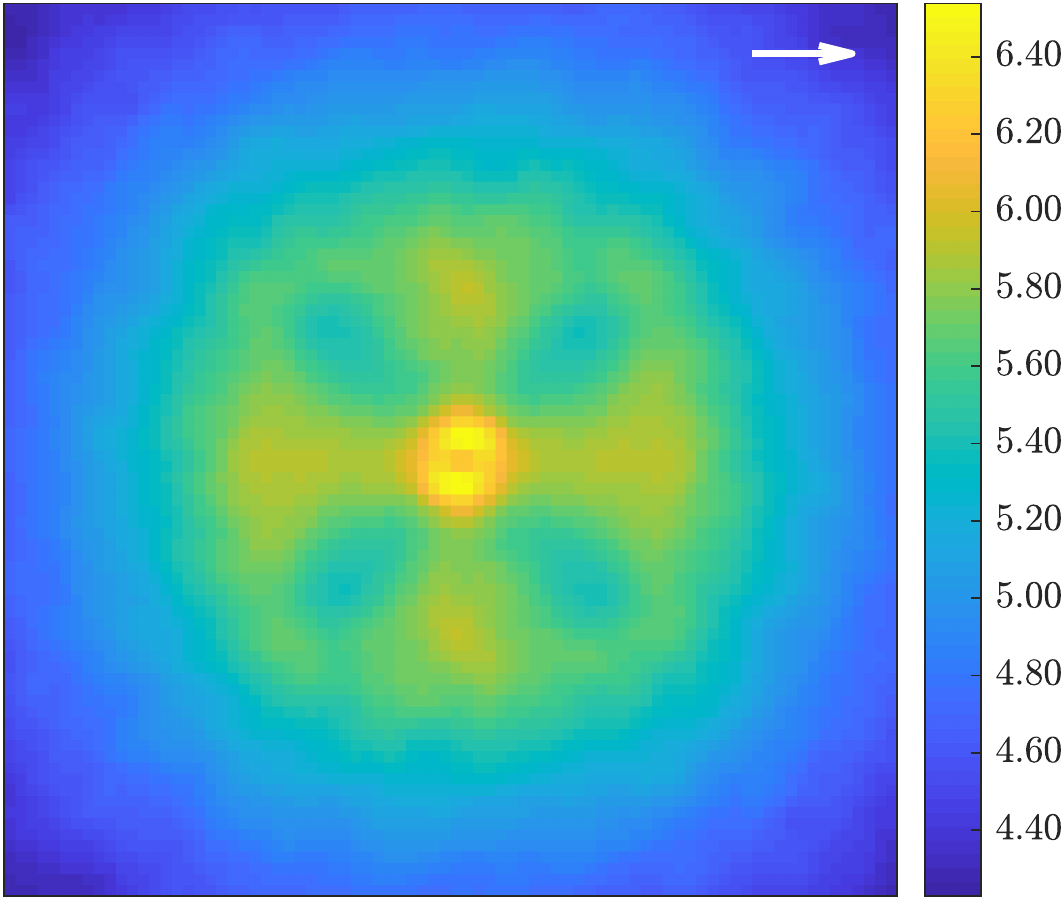}
     \end{subfigure}
    ~
     \begin{subfigure}[b]{0.32\textwidth}
         \centering
         \caption{LSTM PSD (86.1\% Strehl Ratio)} \vspace{-1.5mm}
         \includegraphics[width=\textwidth]{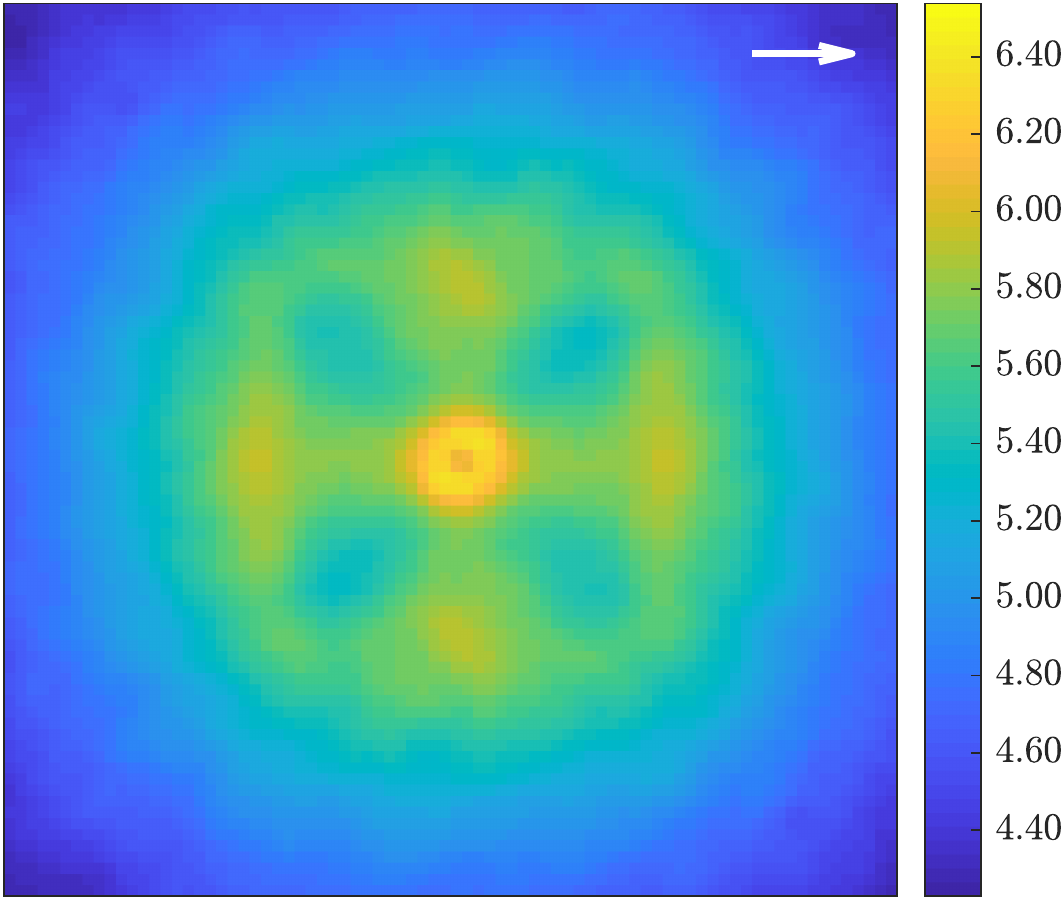}
     \end{subfigure}
     \\
     \vspace{2.5mm}
    \begin{subfigure}[b]{0.32\textwidth}
         \centering
         \caption{Classical/Dense Ratio Image} \vspace{-1.5mm}
         \includegraphics[width=\textwidth]{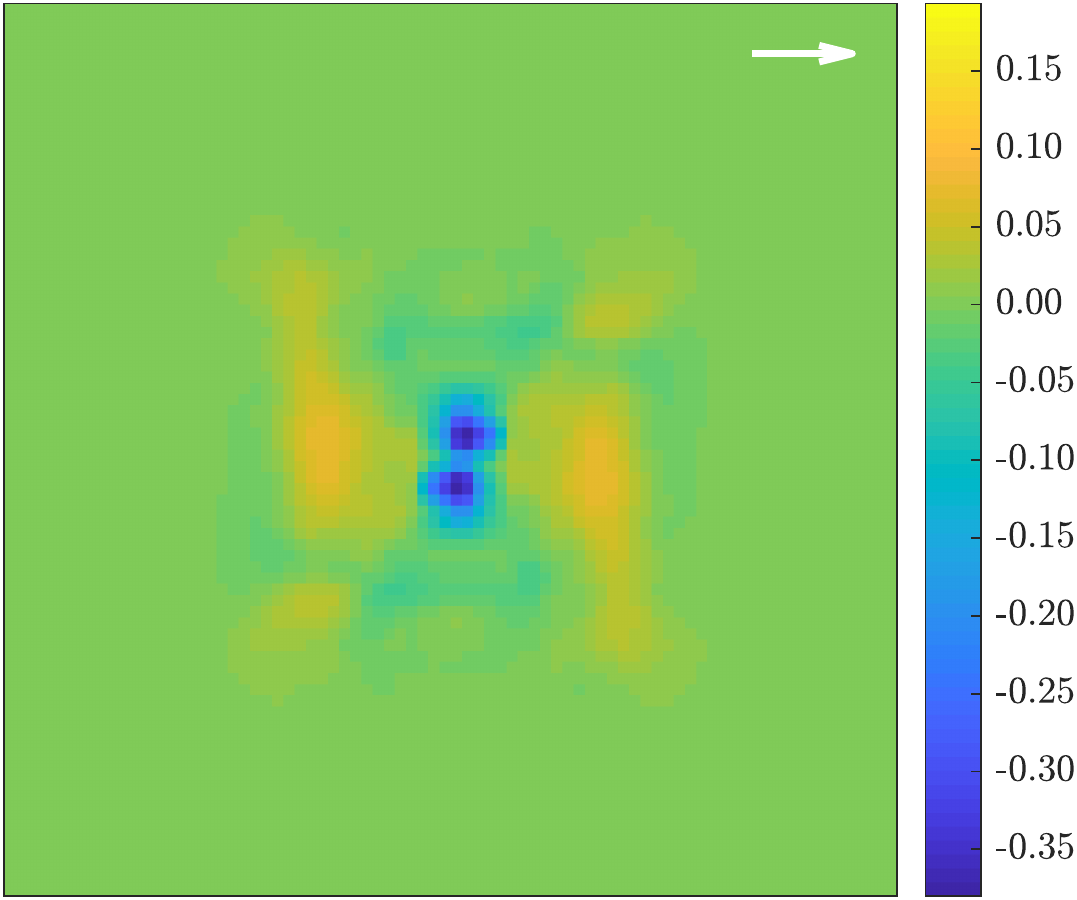}
    \end{subfigure}
     ~
     \begin{subfigure}[b]{0.32\textwidth}
         \centering
         \caption{Classical/LSTM Ratio Image} \vspace{-1.5mm}
         \includegraphics[width=\textwidth]{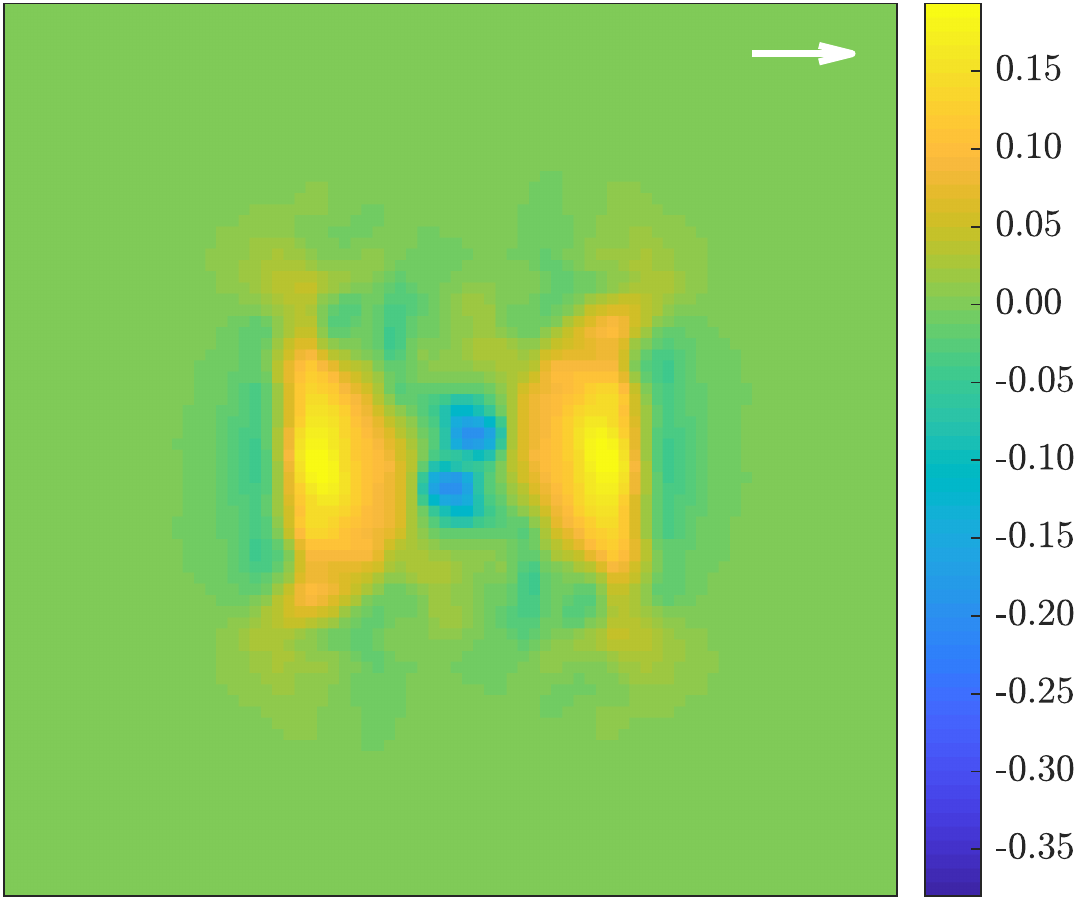}
     \end{subfigure}
    ~
     \begin{subfigure}[b]{0.32\textwidth}
         \centering
         \caption{Dense/LSTM Ratio Image} \vspace{-1.5mm}
         \includegraphics[width=\textwidth]{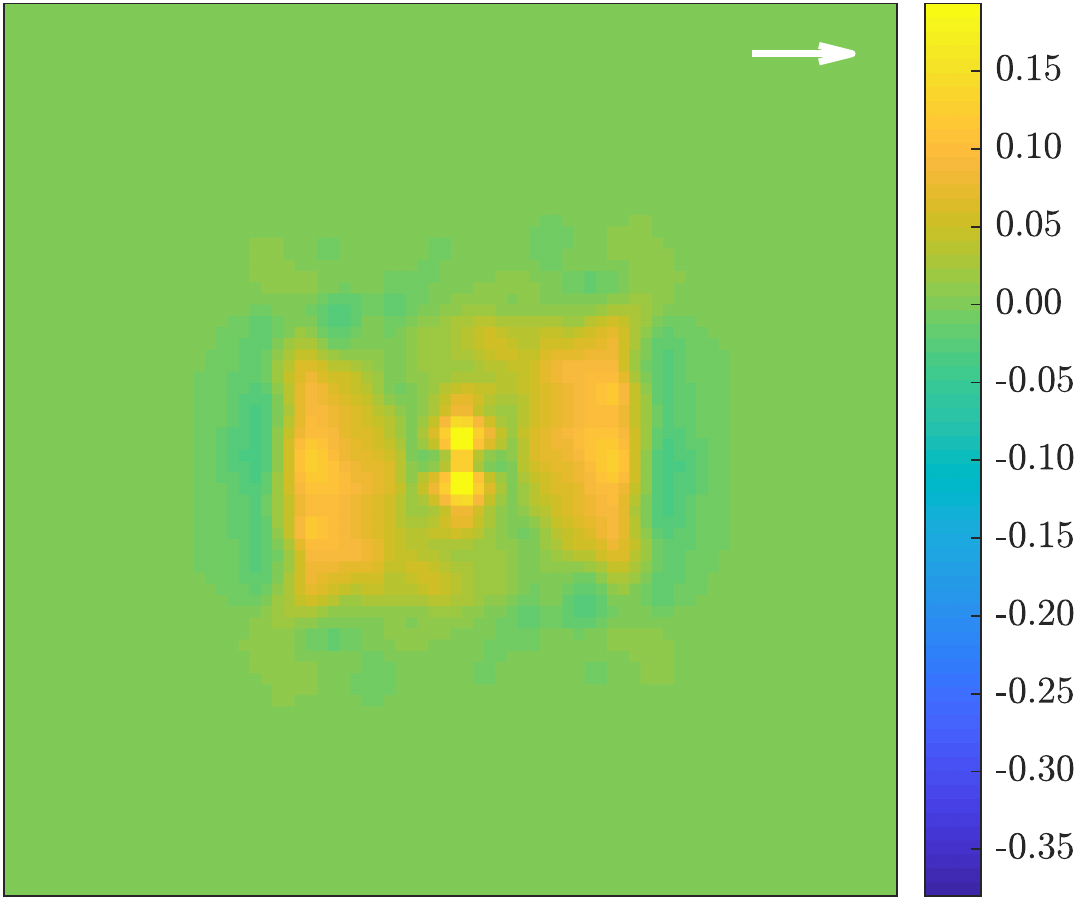}
    \end{subfigure}
     \\
     \vspace{2.5mm}
           \begin{subfigure}[b]{0.32\textwidth}
         \centering
         \caption{Classical/Dense Ratio Image} \vspace{-1.5mm}
         \includegraphics[width=\textwidth]{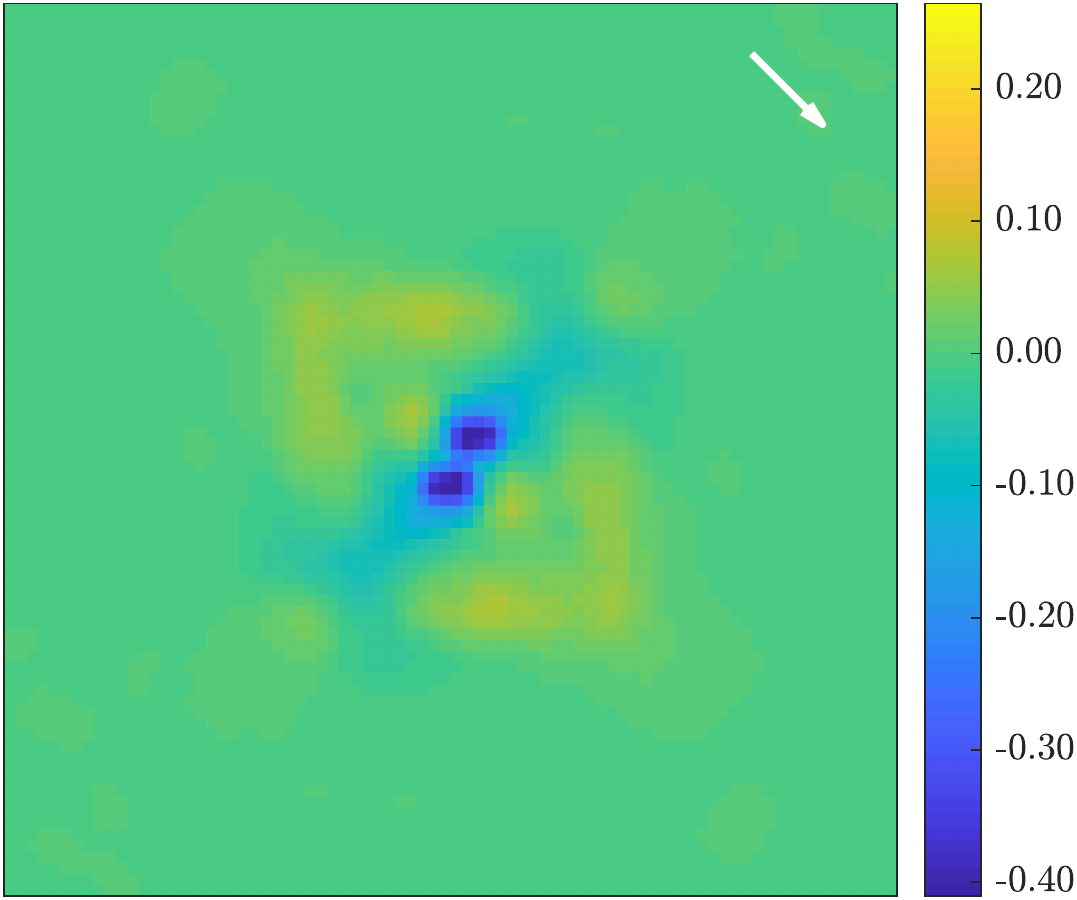}
     \end{subfigure}
     ~
     \begin{subfigure}[b]{0.32\textwidth}
         \centering
         \caption{Classical/LSTM Ratio Image} \vspace{-1.5mm}
         \includegraphics[width=\textwidth]{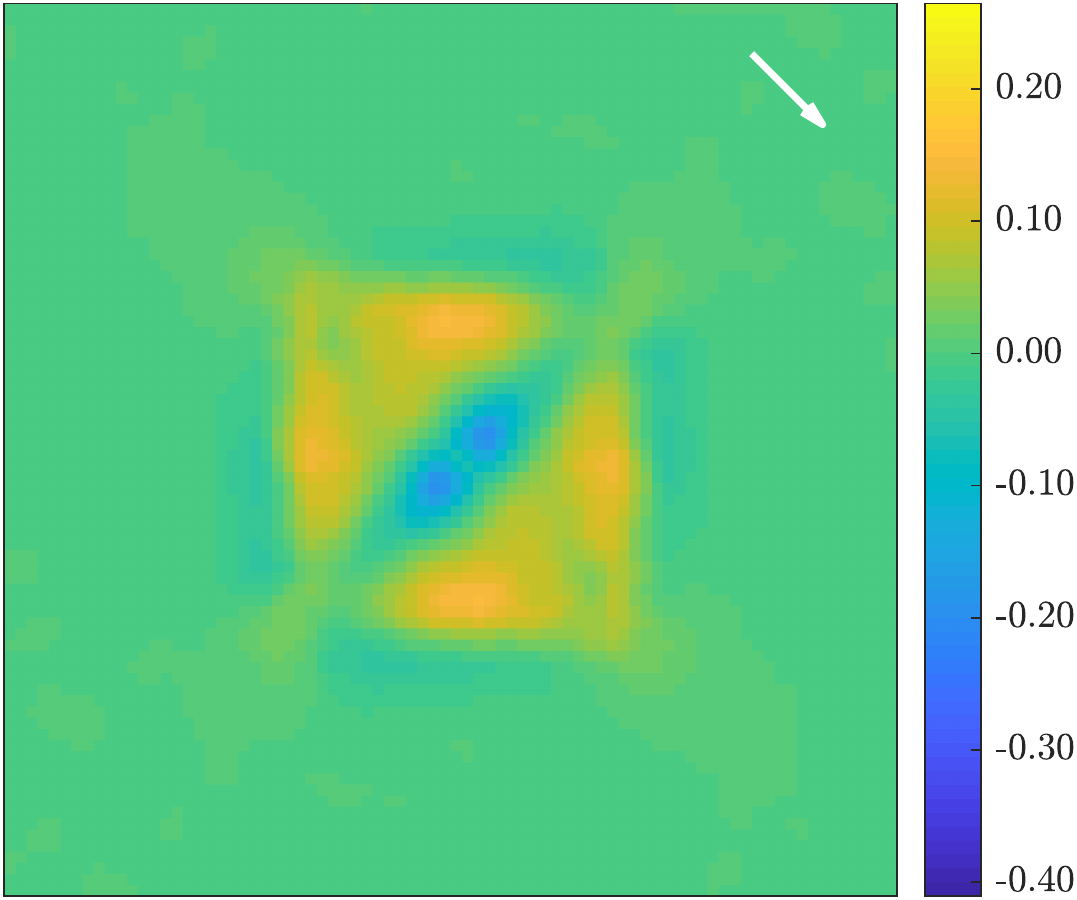}
     \end{subfigure}
    ~
     \begin{subfigure}[b]{0.32\textwidth}
         \centering
         \caption{Dense/LSTM Ratio Image} \vspace{-1.5mm}
         \includegraphics[width=\textwidth]{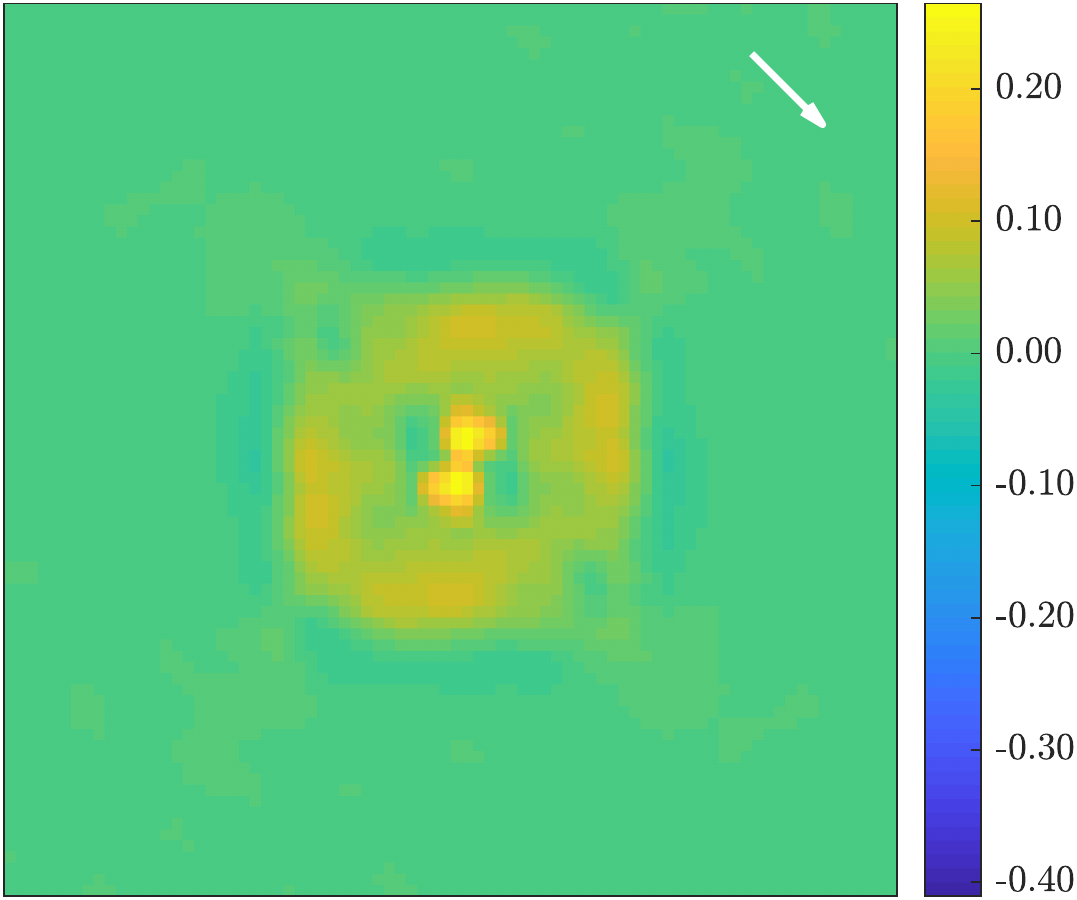}
    \end{subfigure}
    
        \caption{\textbf{Magnitude 8 Residual Wavefront Frequency Analysis:} Example log-scale power spectral density (PSD) plots for wind direction at 0\textdegree~(top) and PSD ratio images of the residual wavefronts for wind directions at 0\textdegree~(middle) and 45\textdegree~(bottom). Our networks clearly show improvements along the direction of the wind in both cases. Although lower frequency performance corresponding to tip and tilt are not as well compensated, these frequencies lie within the typical cutoff radius of a coronagraph and thus will have negligible impact on the system's performance in such settings.}
        \label{fig:psd_plots_8mag}
\end{figure*}

\begin{figure*}
\centering
\captionsetup[subfigure]{labelformat=empty}
      \begin{subfigure}[b]{0.32\textwidth}
         \centering
         \caption{Classical PSD (42.4\% Strehl Ratio)} \vspace{-1.5mm}
         \includegraphics[width=\textwidth]{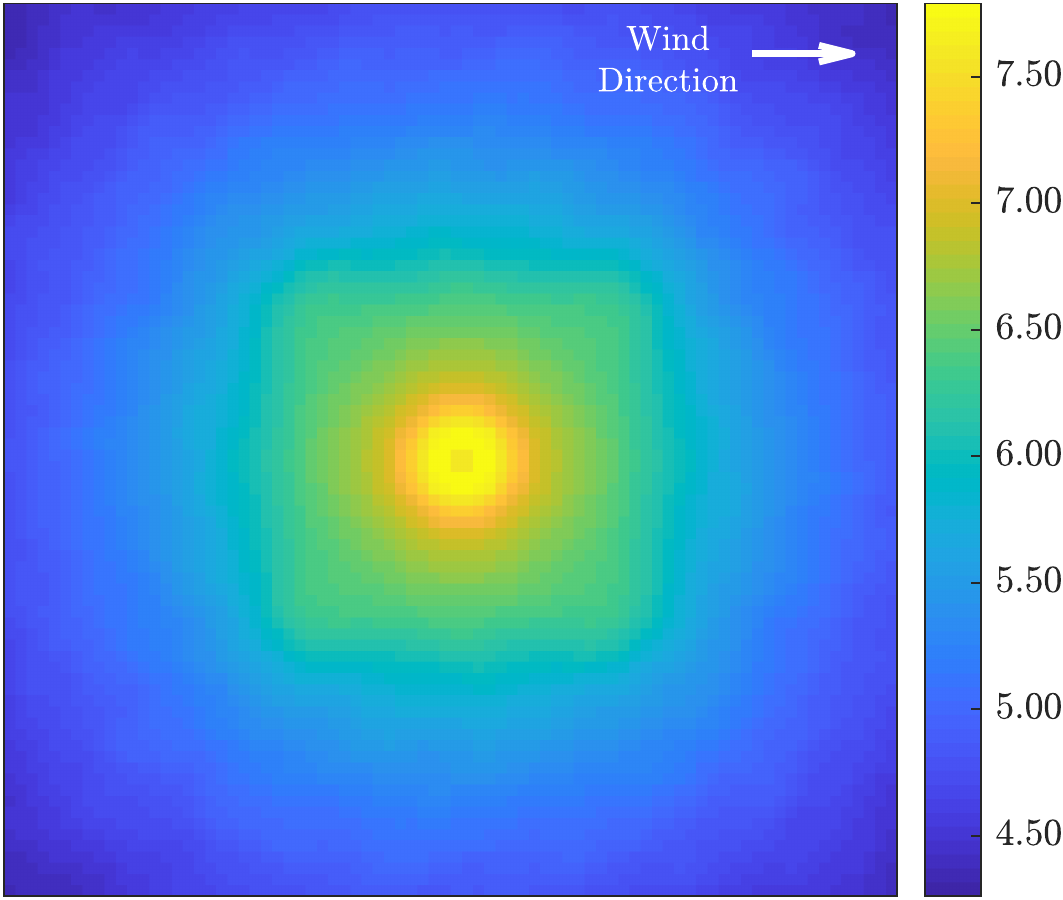}
     \end{subfigure}
     ~
     \begin{subfigure}[b]{0.32\textwidth}
         \centering
         \caption{Dense PSD (65.5\% Strehl Ratio)} \vspace{-1.5mm}
         \includegraphics[width=\textwidth]{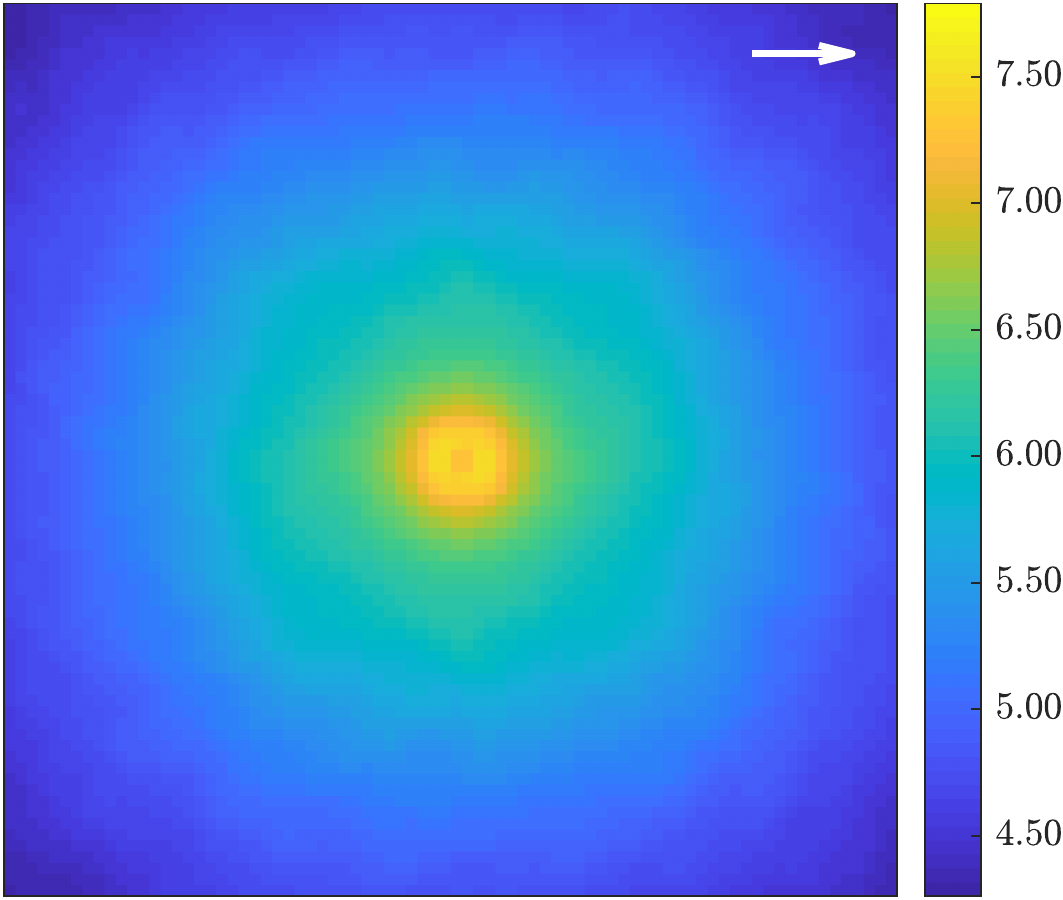}
     \end{subfigure}
    ~
     \begin{subfigure}[b]{0.32\textwidth}
         \centering
         \caption{LSTM PSD (63.0\% Strehl Ratio)} \vspace{-1.5mm}
         \includegraphics[width=\textwidth]{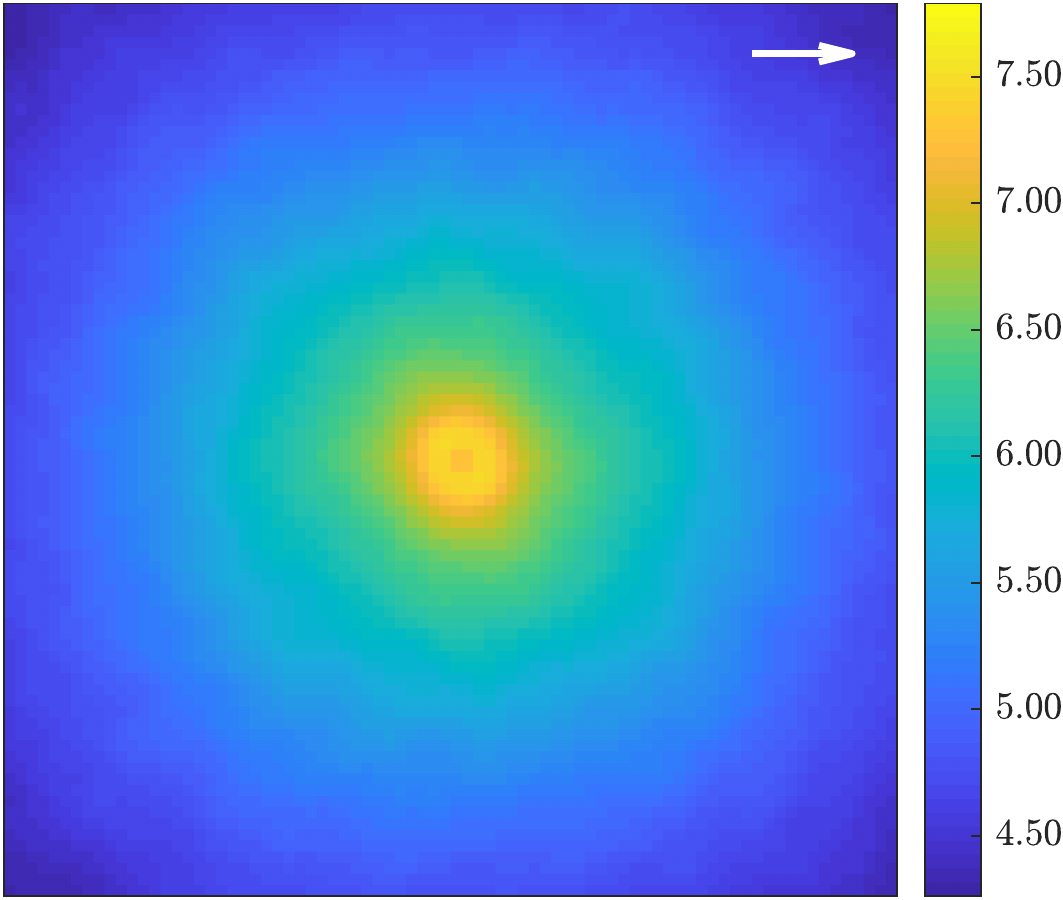}
     \end{subfigure}
     \\
     \vspace{2.5mm}
           \begin{subfigure}[b]{0.32\textwidth}
         \centering
         \caption{Classical/Dense Ratio Image} \vspace{-1.5mm}
         \includegraphics[width=\textwidth]{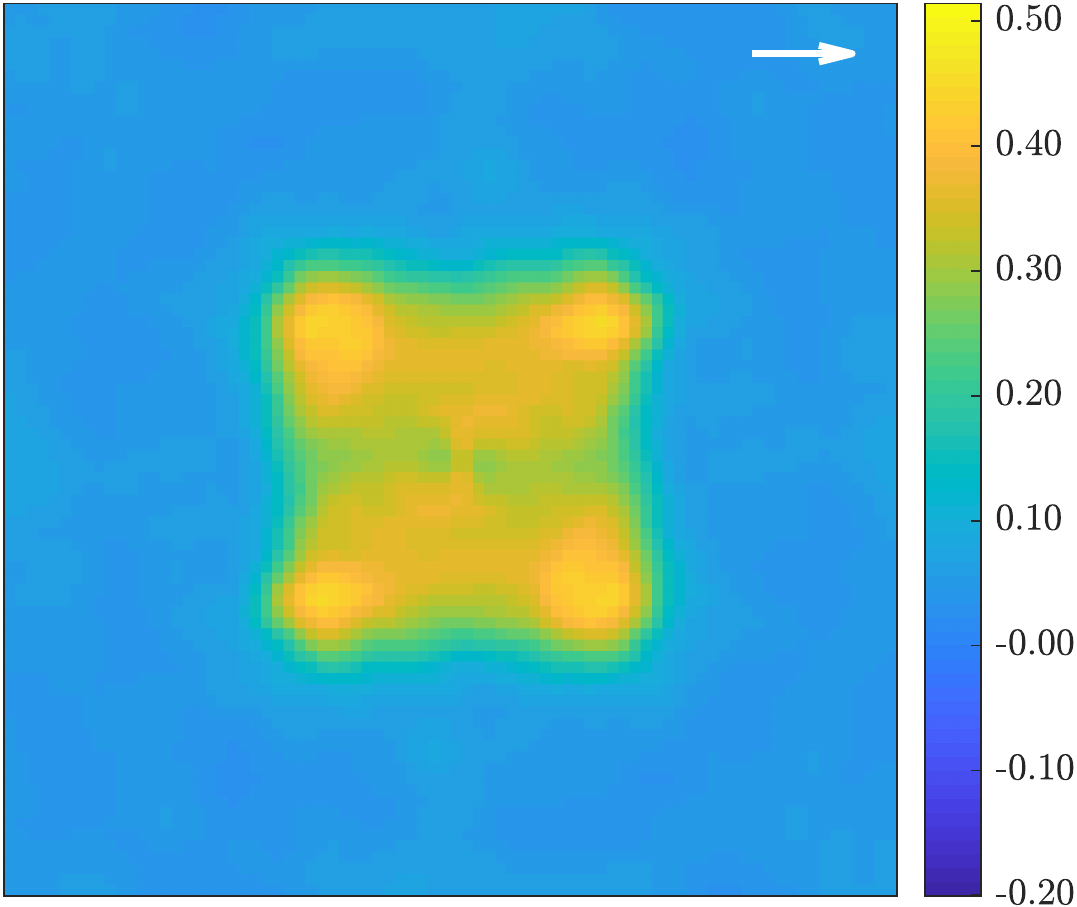}
     \end{subfigure}
     ~
     \begin{subfigure}[b]{0.32\textwidth}
         \centering
         \caption{Classical/LSTM Ratio Image} \vspace{-1.5mm}
         \includegraphics[width=\textwidth]{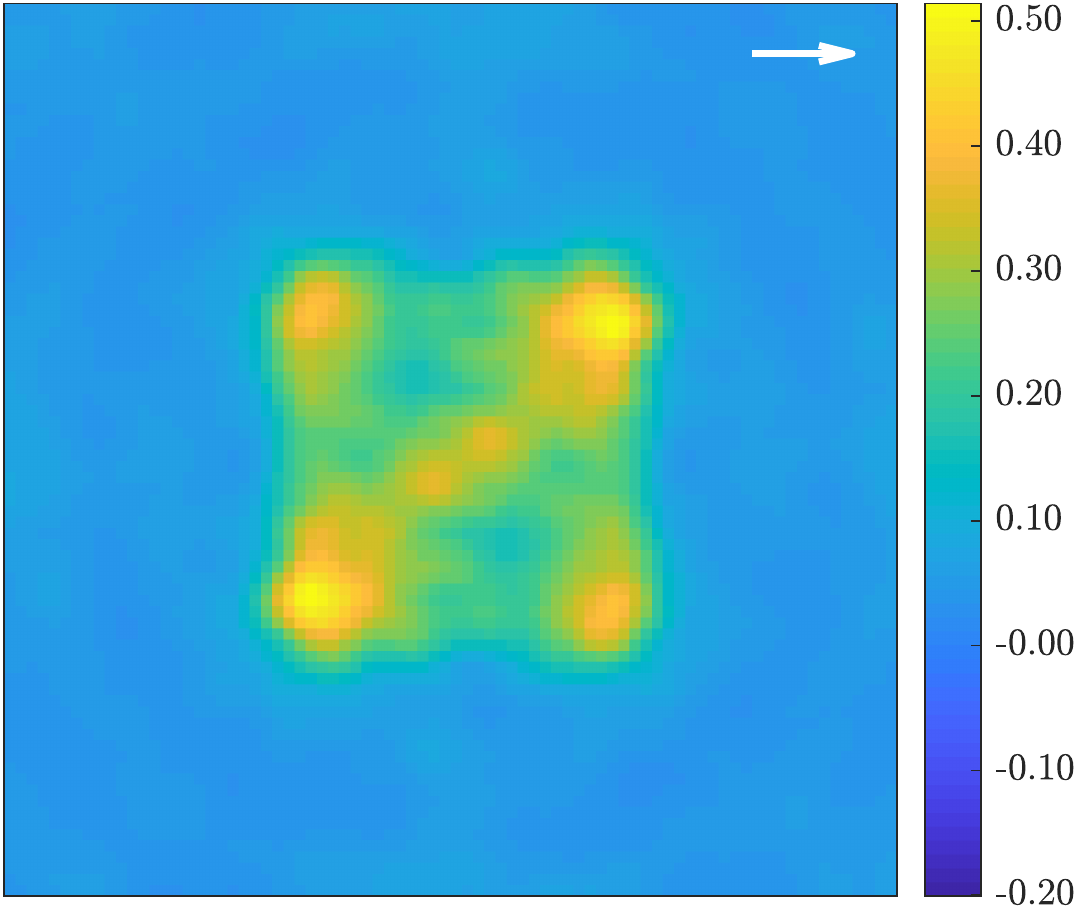}
     \end{subfigure}
    ~
     \begin{subfigure}[b]{0.32\textwidth}
         \centering
         \caption{Dense/LSTM Ratio Image} \vspace{-1.5mm}
         \includegraphics[width=\textwidth]{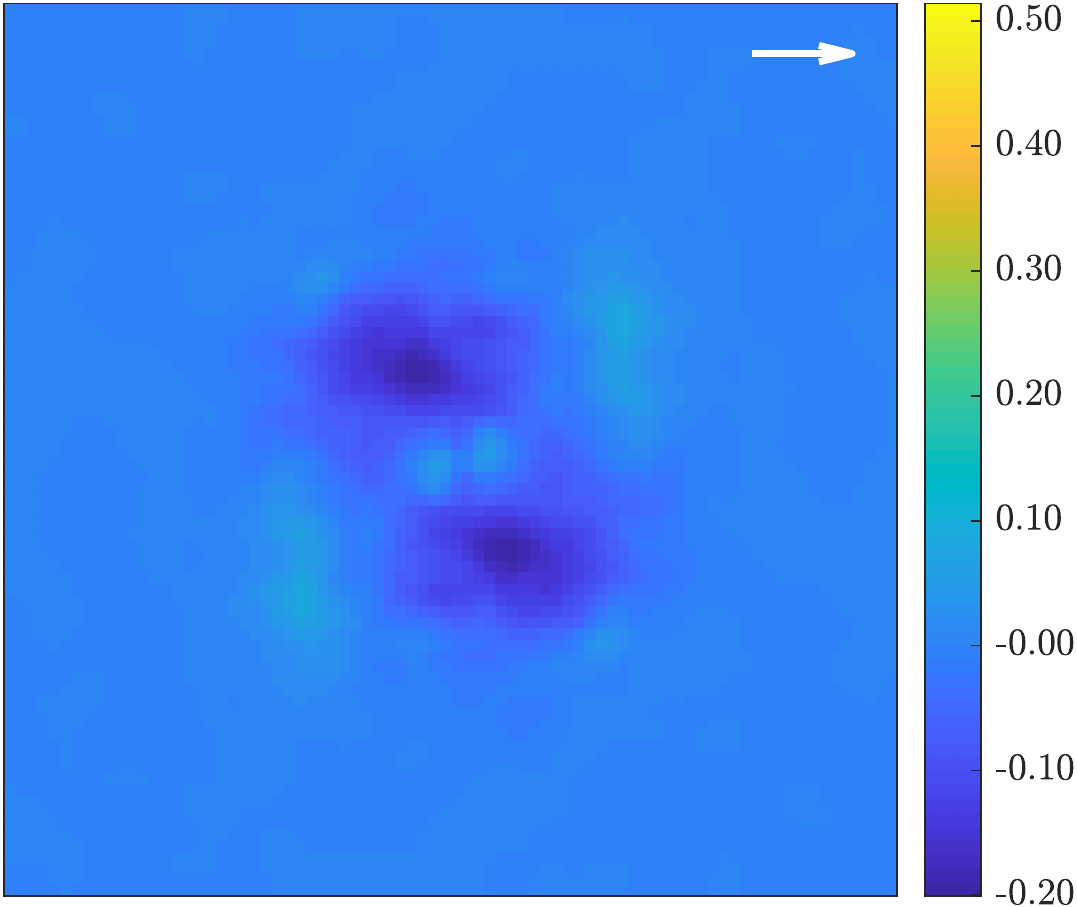}
    \end{subfigure}
     \\
     \vspace{2.5mm}
           \begin{subfigure}[b]{0.32\textwidth}
         \centering
         \caption{Classical/Dense Ratio Image} \vspace{-1.5mm}
         \includegraphics[width=\textwidth]{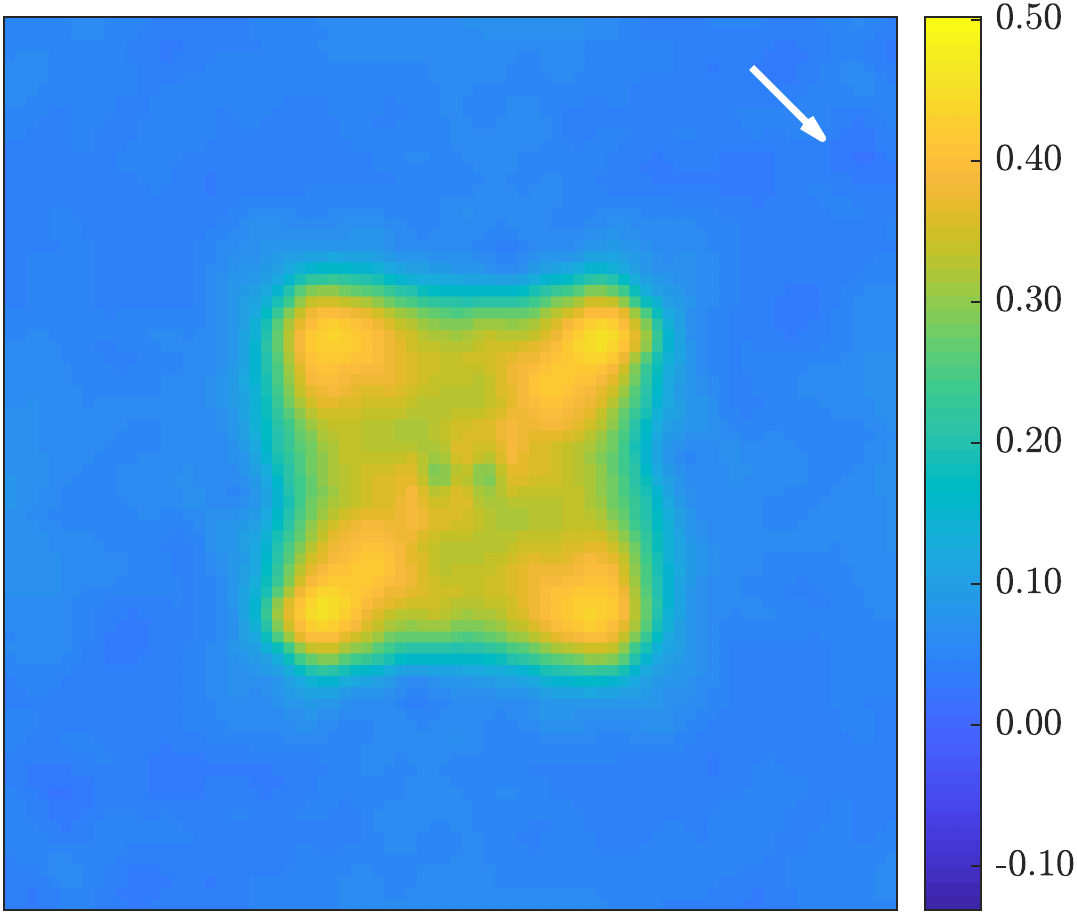}
     \end{subfigure}
     ~
     \begin{subfigure}[b]{0.32\textwidth}
         \centering
         \caption{Classical/LSTM Ratio Image} \vspace{-1.5mm}
         \includegraphics[width=\textwidth]{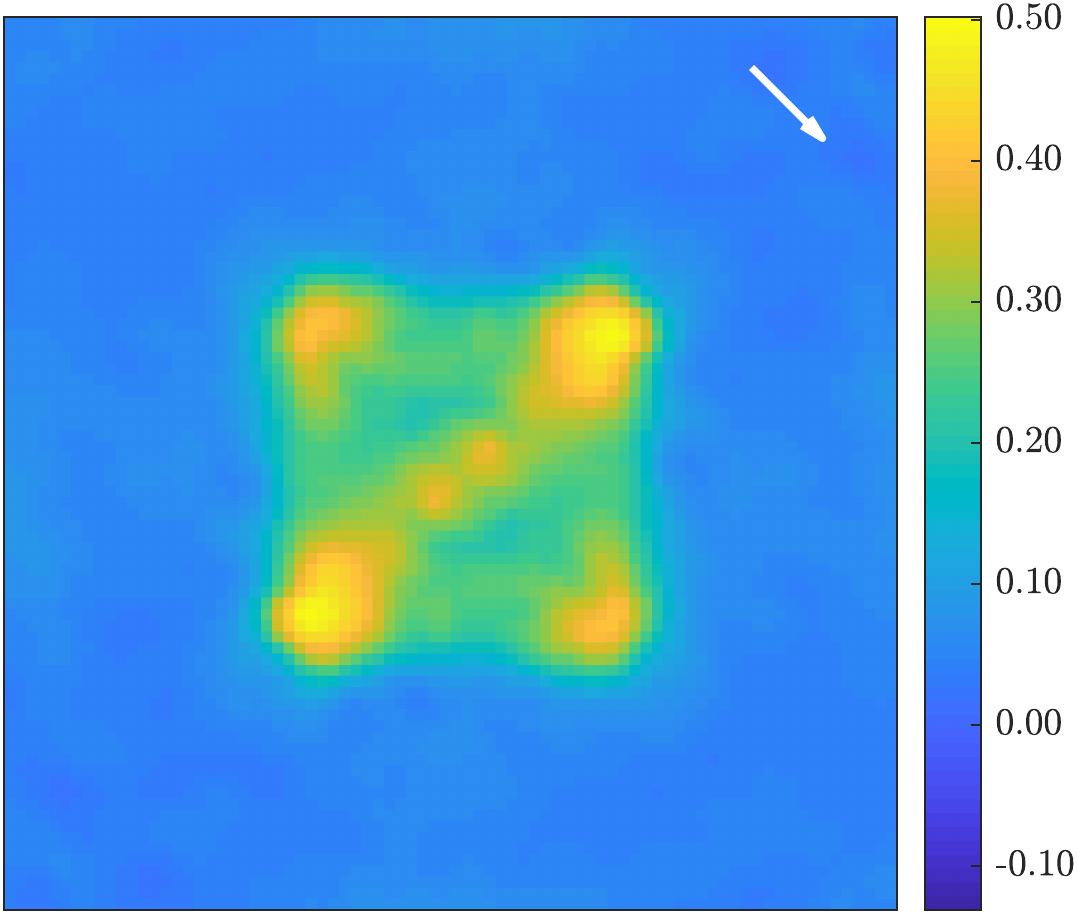}
     \end{subfigure}
    ~
     \begin{subfigure}[b]{0.32\textwidth}
         \centering
         \caption{Dense/LSTM Ratio Image} \vspace{-1.5mm}
         \includegraphics[width=\textwidth]{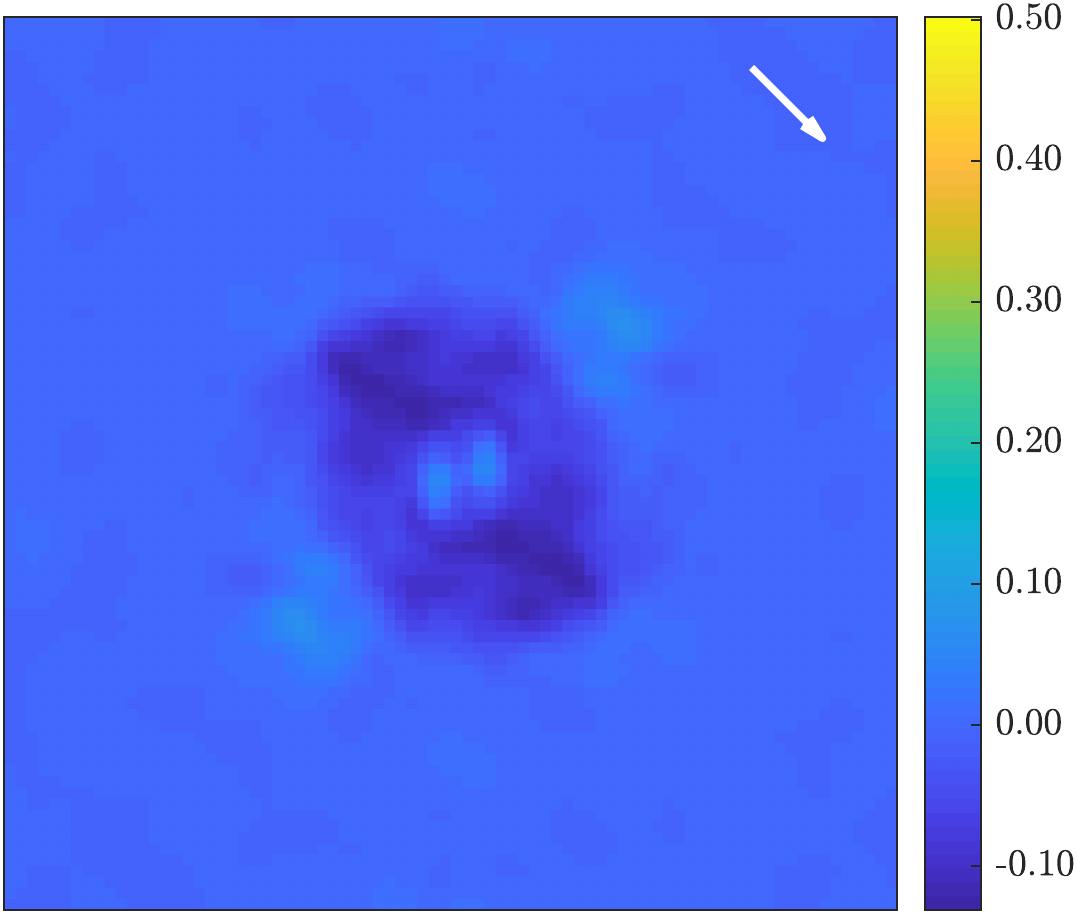}
    \end{subfigure}
    
        \caption{\textbf{Magnitude 10 Residual Wavefront Frequency Analysis:} Example log-scale power spectral density (PSD) plots for wind direction at 0\textdegree~(top) and PSD ratio images of the residual wavefronts for wind directions at 0\textdegree~(middle) and 45\textdegree~(bottom). Our networks clearly show improvements along the direction of the wind in both cases across the entire control radius.}
        \label{fig:psd_plots_16mag}
\end{figure*}

\subsection{Robustness to New Parameters}

While our models were trained over a wide range of atmospheric conditions, some variables remained fixed throughout training and testing. Here we explore the robustness of our previously trained models when these unseen conditions are met for the first time.

Figure~\ref{fig:results_wind_change_run} shows an example simulation, where the wind changes drastically throughout the simulation. Here the wind conditions are changed every 50 time steps, while in training our networks only experienced more stable conditions. Our networks are still able to reliably outperform the classical integrator.

To see how well our networks performed when the atmospheric layer distribution changed we tested our models (trained on a 3 layer atmosphere) under a 6-layer atmosphere. Figure~\ref{fig:results_more_atm} shows the results for a single simulation for a magnitude 8 and 16 NGS. Here we see that while the dense network has no difficulty generalizing to additional layers, the LSTM network performance somewhat suffers while the classical method remains unaffected. This result is not entirely unexpected as the number of atmospheric layers may greatly increase the complexity of the required predictions. This points towards the need to train on richer atmospheres, potentially with more layers, or to better understand the expected atmospheric conditions.

Finally in Figure~\ref{fig:results_change_L0} we vary the outer scale of the atmosphere (L0) and show the average strehl ratio over five simulations for each value. Although our training data had a constant L0 value of 30m, we see almost no change in performance for all three methods as this value varies.

\subsection{Frequency Analysis}

To better understand the performance gains of our approach, we investigate the average frequency content of the residual atmospheric wavefront over the course of a simulation. At each time step we calculate the power spectral density (PSD) of the residual wavefront, applying a Hamming filter to account for high frequencies caused by the telescope pupil. This is then averaged over the entire simulation. Because the residual PSD of an AO system is a good approximation of an ideal coronagraphic image (\cite{correia2020performance}), any improvements found in the residual PSD of our methods should result in improved performance of a coronagraphic image. To reduce the noise in the PSD we increased the simulation duration by a factor of twenty. We also fixed the wind direction for all layers to the same horizontal or diagonal direction to aid with our interpretation of the results.

Figures~\ref{fig:psd_plots_8mag} and ~\ref{fig:psd_plots_16mag} show example comparison plots for extended simulations over 50 seconds with NGS magnitude of 8 and 16. In both cases the top row shows the PSD result in log scale when the wind is blowing in the horizontal direction for all three methods, cropped to better show the AO control radius. These figures indicate the spatial frequencies where the residual wavefront was not well-corrected. The ratio maps, shown in the middle (for horizontal wind) and bottom rows (for diagonal wind directions), are the ratio between the two listed methods. These ratio images help give a clearer picture of the frequencies where each method performs best.

While the PSDs in the top row of Figure~\ref{fig:psd_plots_8mag} may appear similar for the magnitude 8 case, it is clear that the central core is elongated for the classical method. This indicates reduced compensation of servo-lag error. Furthermore, the ratio images for the horizontal and diagonal wind directions suggest both our networks improve performance along the wind direction while sacrificing some low-frequency performance near the core of the PSD. For the magnitude 16 case shown in Figure~\ref{fig:psd_plots_16mag} our gains are much stronger. The ratio images show improvements across the entire control radius.

Our results suggest that while both networks are able to improve upon the classical method, the LSTM network may be better suited for use in low-magnitude star scenarios where servo-lag dominates while the dense network is better suited for faint stars where noise is the dominant source of error.

\section{Conclusions}

\subsection{Summary}

In this paper we presented a novel method for supervised training of predictive closed loop adaptive optics controllers. Our models improve on classical methods for both low-light and high-contrast scenarios, increase robustness to a wide range of seeing conditions, and demonstrate predictive capabilities. We find that our LSTM network may be better suited for use in low-magnitude NGS scenarios where servo-lag dominates while the dense network is better suited for faint stars where noise is the dominating source of error. While there is still work to be done before on-sky testing can be done, we present both promising simulated results and a clear path towards these goals.

\subsection{Hardware Implementation}

While we have validated our method in simulation, the next step is to show bench results to verify our claims. As is often the case when moving from simulation to hardware (and eventually on-sky), many issues may arise due to underlying differences in the data. We do not foresee substantial issues beyond potentially acquiring additional training data from our hardware.

Real-time application on an actual telescope will require more work to incorporate into existing AO pipelines. Due to the high speeds at which these pipelines operate, achieving real-time operation will be a challenge. Our dense model currently runs at 200Hz without any GPU optimization. In contrast, convolutional LSTM methods are not widely supported by optimization tools so additional low-level GPU accelerated code may be required to achieve real-time performance. Fortunately, there is a growing body of work dedicated to optimizing trained networks (\cite{he2017channel}), and custom hardware for  high-speed inference (\cite{lacey2016deep}). Additional changes to our networks to fit a run-time budget are also possible (i.e., adjusting the number of layers, filters, or dense blocks).

\section*{Acknowledgements}

The Dunlap Institute is funded through an endowment established by the David Dunlap family and the University of Toronto. The University of Toronto operates on the traditional land of the Huron-Wendat, the Seneca, and most recently, the Mississaugas of the Credit River; the author is grateful to have the opportunity to work on this land. R. Swanson and K. Kutulakos gratefully acknowledge the support of the Vector Institute and NSERC under the RGPIN, COHESA and RTI programs. S. Sivanandam is supported by the NSERC Discovery Grant program C. Correia is supported by FCT grant UIDB/00099/2020. Special thank you to Daniel Delidjakov for graphic design inspiration and support.

\section*{Data Availability}
The source code and data underlying this article will be shared on reasonable request to the corresponding author.




\bibliographystyle{mnras}
\bibliography{refs} 







\bsp	
\label{lastpage}
\end{document}